 \documentclass[aps,showpacs,twocolumn] {revtex4}
\usepackage{graphicx}
\usepackage{color}

\begin{document}

\title{On regular and random two-dimensional packing of crosses}

\author{Ralf Stannarius and Jonas Schulze}

\affiliation{
 Otto-von-Guericke-University Magdeburg, Institute of Physics,
Universit\"atsplatz 2, D-39106 Magdeburg, Germany
}

\begin{abstract}
Packing problems, even of objects with regular geometries, are in general non-trivial.
For few special shapes, the features of crystalline
as well as random, irregular two-dimensional (2D) packings are known. The packing of 2D crosses does not yet
belong to the category of solved problems.
We demonstrate in experiments with crosses of different aspect ratios (arm width to length) which packing fractions are actually achieved
by random packing, and we compare them to densest regular packing structures. We determine local correlations
of the orientations and positions after ensembles of randomly placed crosses were compacted in the plane until they jam.
Short-range orientational order is found over 2 to 3 cross lengths. Similarly, correlations in the spatial distributions of
neighbors extend over 2 to 3 crosses. Apparently, there is no simple relation between the geometries of the crosses and peaks
in the spatial correlation functions. Some features of the orientational correlations, however, are intuitively evident.
 \end{abstract}


\maketitle

\section{Introduction}
Prediction of packing configurations of identical particles is in general a tough problem, even for very simple particle geometries and without spatial restrictions.
For example, the densest regular three-dimensional (3D) packing of spheres with space filling of $\pi/\sqrt{18}\approx 74\%$ was already conjectured by Johannes Kepler,
more than four centuries ago. But irrespective of the apparent simplicity of this problem, it took until the beginning of 21$^{\rm st}$ century to find the proof for this conjecture \cite{Hales2005}.
Spatial restrictions like enclosure in cylindrical, cuboid or other containers add even more complexity
\cite{McGeary1963,Mughal2012,Reimann2013,Harth2015,Levay2018}.

When the spheres are not monodisperse, they can be packed more efficiently \cite{Baranau2014}.
Besides spheres, other particle shapes as for example ellipsoids \cite{Donev2004}, polyhedra \cite{Torquato2009,Haji2013,Teich2016} or spherocylinders \cite{Kyrylyuk2011b} have attracted the interest of researchers. A number of similar studies on packing problems in 3D included also non-convex particles
like stars or hexapods \cite{Zhao2016,Zhao2020}. The practical relevance of the latter type of objects may be found in construction industry and architecture
\cite{Dierichs2015,Keller2015}, when one is looking for a material that forms stable aggregates, but on the other hand can be disassembled again with little efforts, to redesign its global shape.
An excellent and comprehensive review of 3D packing problems of particles of various shapes has been published by Torquato and Stillinger \cite{Torquato2010}.
%

Often, when a set of randomly placed objects is compacted, a regular crystalline ground state is not reached.
Instead, more or less random local arrangements are obtained both in experiments and simulations.
For example, pouring hard objects into a container does not lead to the spontaneous formation of a regular, crystalline configuration. Even shaking or shearing ensembles of spheres leaves them in arrangements where packing is random, and the packing fraction of around $\phi_{\rm rcp}~\approx 64\%$ remains well below the optimum value for the crystalline configurations
\cite{Knight1995,Clement1998,Richard2005,Anikeenko2007,Kapfer2012,Rietz2018}.
This jammed state is irregular and in general much less densely packed than the optimum space-filling arrangement of the particles.

Altogether, we have to distinguish strictly three different packing problems:
The first one is the question of optimal packing, in an infinitely extended volume or under particular geometrical constraints (e.~g.~\cite{Atkinson2012,Haji2013,Mughal2012,Harth2015,Levay2018,Bautista2018,Meng2020}).
For convex centro\-symmetric particles, densest packing structures are crystalline \cite{Atkinson2012}.
In general, however, this state is not reached after compaction of a random ensemble of particles.

The second problem is the description of externally agitated matter, for example granular ensembles under mechanical vibrations (e.~g.~\cite{Macrae1957,Nowak1997,Remond2006,Rosato2010,Sanchez2015,Walsh2016,Windows-Yule2017}). Such a system in a non-perfectly ordered state is able to explore the configuration space. Here, the entropy of the individual configurations
comes into play.
In simulations, one can force the system by repeated growth and shrinking of the ensemble
\cite{Donev2006}, or by adding Brownian motion \cite{Wojciechowski2004,Martinez2013,Cadotte2016,Hou2020}.
In many respects, these systems can be considered as analogues of
colloidal crystals. Container vibrations play the role of thermal fluctuations. The individual particles can
find some more compact configurations. In some of these systems, local order may
develop spontaneously.
For vibrated rods in quasi-2D monolayers, for example, a tetratic local orientational order was discovered in experiments
\cite{Sanchez2015}.
Tetratic order was also reported in simulations of ensembles of densely packed rectangles \cite{Donev2006},
Similar investigations were performed for other shapes such as kite-shaped  \cite{Hou2020}, 
bow-shaped \cite{Martinez2013},
or square \cite{Wojciechowski2004,Walsh2016} 
particles.

The third problem is the understanding of the jammed state which is reached in an athermal system
when the packing fraction is gradually increased by compaction. When the jamming transition has occurred,
the system is in equilibrium. No further dynamics is considered. One important parameter is the space filling $\phi$
(area filling in 2D) at the jamming point. In the present study, we will consider the third problem.
We search for the packing density and structure that characterizes the onset of jamming.

In two dimensions (2D), packing problems are simpler than in 3D but nevertheless far from being generally understood.
A proof of the densest (hexagonal) disk packing with area filling of $\phi=\pi/\sqrt{12}\approx 0.9069$ was published
by Fejes \cite{Fejes1972}. Since the hexagonal lattice optimizes 2D packing locally and globally, filling a plane with disks yields an almost crystalline structure with few defects. These will gradually disappear when the system is appropriately agitated.

When moving from disks to more complex shapes like polygons
\cite{Xu2017,Zhao2019}, ellipses and dimers \cite{Schreck2010}, complex packing structures are found.
Non-convex shapes composed of straight or bent segments have been considered in numerical simulations
e.~g. in Refs. \cite{Zheng2017,Marschall2020,Meng2020}. 
In contrast to the number of numerical studies, experimental work has been performed only scarcely.

\begin{figure}[htbp]
\center
\includegraphics[width=0.45\columnwidth]{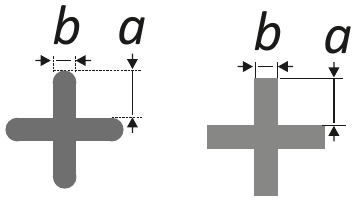}\vspace{3mm}
\caption{\label{fig:crosses1}
Definitions of the arm length $a$ and width $b$ for rounded and simple crosses.
}
\end{figure}

The packing of cross-shaped objects in a plane has been analysed before in different contexts and under different aspects
(e.~g. \cite{Atkinson2012,Zheng2017,Marschall2020,Meng2020}), since it is both interesting from a fundamental point of view and for practical purposes. We refer here to regular flat objects with fourfold
symmetry, a horizontal and four vertical mirror planes ($D_{4h}$). Their extension in the third dimension is
disregarded and we consider only their 2D projections. The two types of crosses considered in this study are shown in Fig.~\ref{fig:crosses1}.
The most relevant geometrical feature of crosses is their arm width to arm length ratio,
which will determine their optimal packing structure (the one with the best area filling). We denote here the length of the arm as $a$, its width as $b$
(see Fig.~\ref{fig:crosses1}),
the aspect ratio as $\rho=b/a$, and the cross 'size' as $\ell=2a+b$ (the distance of the ends of opposing arms). The simplest
crosses have arms with flat ends.
The limiting cases of those structures for $a\rightarrow 0$, $\rho\rightarrow\infty$ are squares and for $b\rightarrow 0$, $\rho\rightarrow 0$ crossed line segments. A slight variation is the cross shape with rounded arms. It has been considered
in some studies in literature.

There has been only little attention to packing experiments with crosses in 2D so far.
The most relevant experimental study in the context of the present work was performed by Zheng et al. \cite{Zheng2017}.
The authors conducted shear experiments with stress-birefringent cross-shaped objects. The major goals of that study were the characterization of the onset of jamming, the distribution of stresses in the ensemble, and the pressure characteristics near the jamming point.
Their objects are nearly equivalent to
crosses with rounded ends with $\rho\approx0.4$ in our notation, except for rounded corners in the center.
Zheng et al. employed uniaxial and biaxial compression. Both
approaches led to slightly different jamming points: under uniaxial compression, the jamming transition was reached at
about 47.5 \% space filling ($\phi=0.475$), while under biaxial compression this value was slightly higher, on average by about 0.25 \%.
A systematic investigation of the random two-dimensional (2D) packing characteristics of such objects has not yet been performed there.

Theoretical studies in literature \cite{Meng2020,Marschall2020} also described crosses with semicircular ends of their arms, which facilitates numerical simulations
We will shortly discuss a few consequences of this specific modification for random packing in Sec.~\ref{sec:disc}.
Marschall and Teitel \cite{Marschall2020} studied shearing
of crosses with twofold and fourfold symmetries in a numerical simulation. In our notation, their cross shapes correspond to an aspect ratio of $\rho = 0.5$. Focus of that paper was on the relation between pressure and strain rate in shear experiments,
but the authors also reported a value of the packing fraction. They found $\phi = 0.66$ for the jamming transition.
The authors also reported numerical results for pair correlation functions, and they computed the tetratic
correlation, separately for neighboring crosses in the direction along the arms and diagonal to them.
We will discuss this in the context of our experimental findings below, in Sec.~\ref{sec:disc}.

The present manuscript is organized as follows:
Before we describe the experiments with random arrangements of the particles, we will first recollect structures that are best fitted candidates for an optimum space filling. This allows us to judge the packing efficiency reached by compaction compared
to the maximum achievable $\phi$ for a given cross shape. Within the limits of $0<\rho<\infty$, very different optimum crystalline arrangements exist \cite{Atkinson2012}.
While our uniaxial compression technique does naturally not result in crystalline packing, some features of these structures will be found in the random arrangements that we study experimentally.

In the second part of this study, we present experiments with plastic crosses of different aspect ratios which are arranged randomly on a table and
then condensed to a jammed cluster that cannot be compressed further without deforming or breaking the objects. The packing achieved that way represents a
good approximation for the random close packing of the respective cross geometry. These irregular arrangements have packing fractions $\phi$ that are considerably smaller than those of
the optimum regular structures, particularly when the aspect ratio $\rho$ is low (thin crosses). We will discuss the short-range spatial correlations
in these arrangements and the mutual orientational correlations of neighbors on the basis of experimental findings.

\section{Crystalline packings}
\label{sec:cryst}

\begin{figure}[htbp]
\centerline{
\includegraphics[width=0.8\columnwidth]{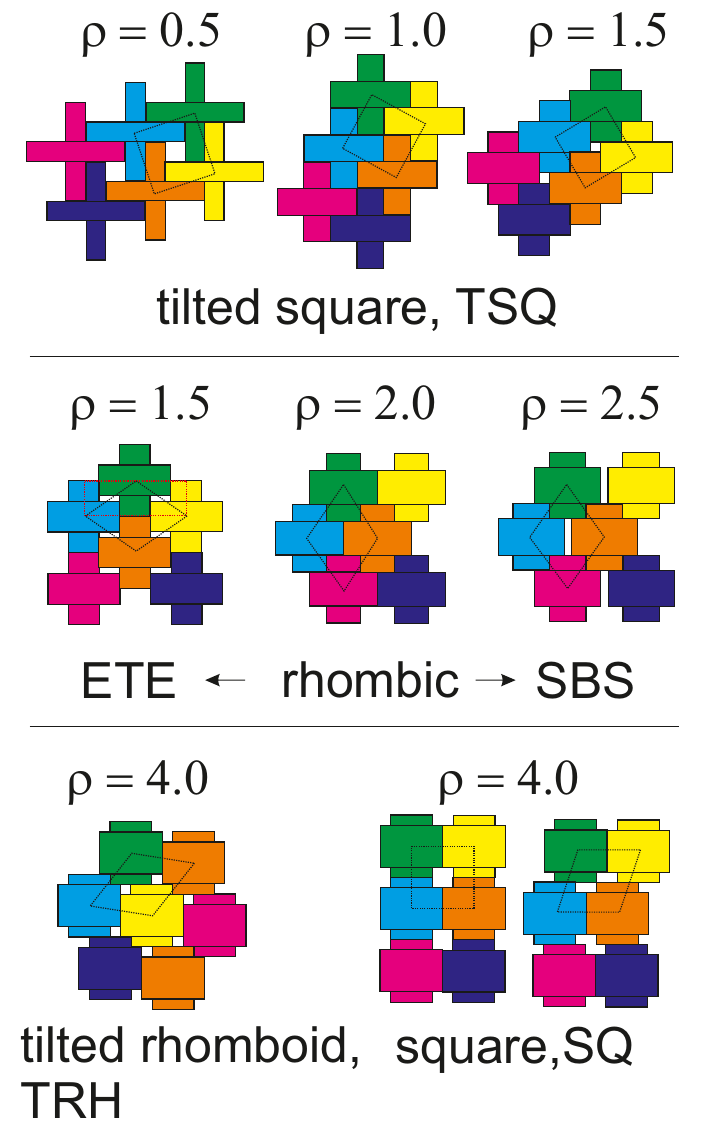}
}
\caption{\label{fig:1}
Regular packings of crosses for selected aspect ratios $\rho=b/a$. The geometric parameters arm length $a$ and arm width $b$ are shown for crosses with
flat and rounded ends, top right.
Tilted square lattices are shown in the top row, rhombic and square lattices in the bottom row.
}
\end{figure}

Before we introduce our experimental results, we recollect the regular crystalline arrangements found to be the best space-filling lattices. The following considerations agree with the results reported in the Supplemental
Material of Ref.~\cite{Atkinson2012}. Atkinson's study considers crosses with straight edges.
While we are not aware of a strict mathematical proof that all the structures described in the following are indeed the optimal arrangements, they
definitely represent lower limits for the optimum packing fraction $\phi_{\max}$.
The optimum packing structures for flat-ended crosses (Fig.~\ref{fig:crosses1} right) are shown in Fig. \ref{fig:1}.

\begin{figure}[htbp]
\centerline{
a)\includegraphics[width=0.5\textwidth]{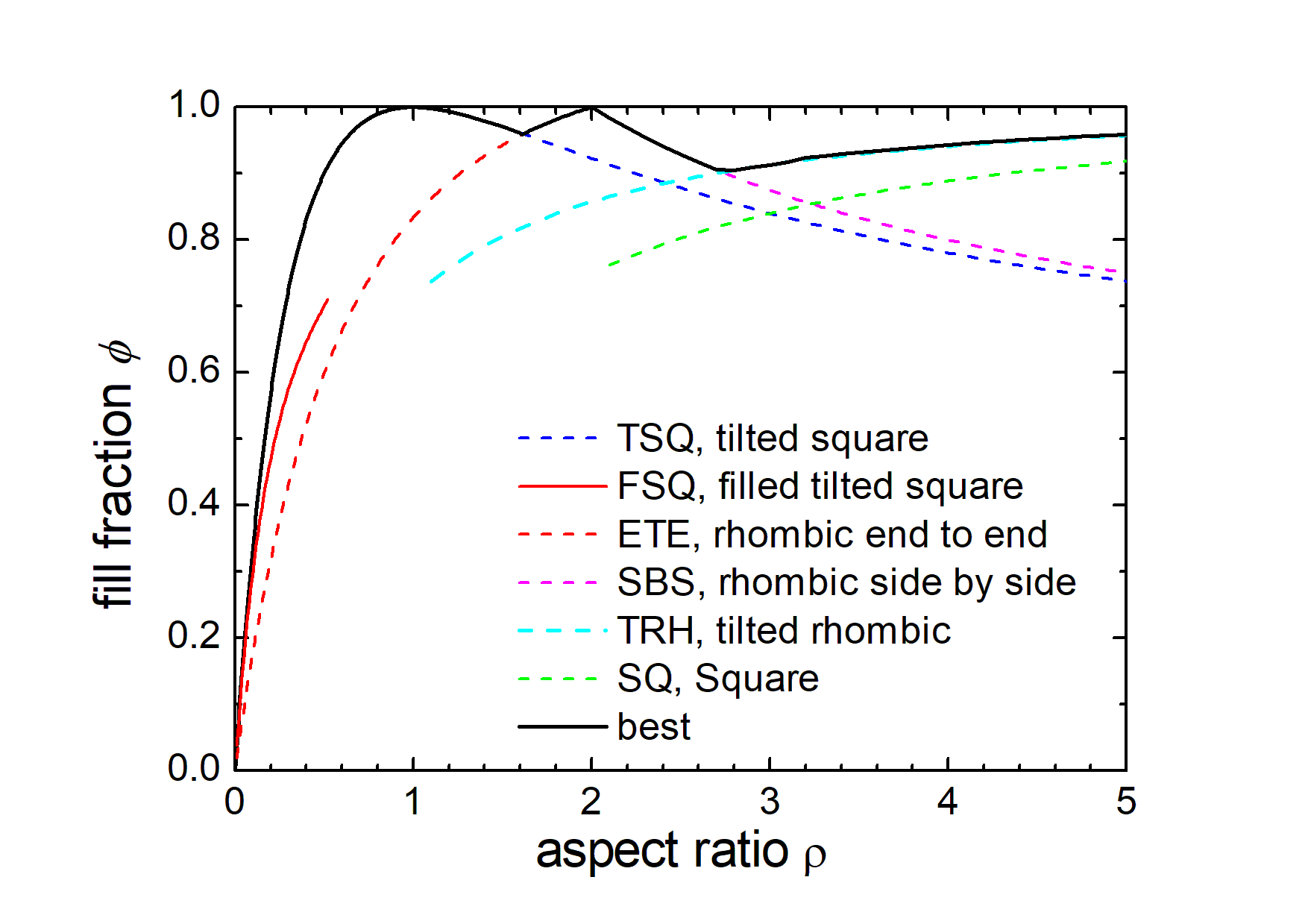}}
\centerline{
b)\includegraphics[width=0.5\textwidth]{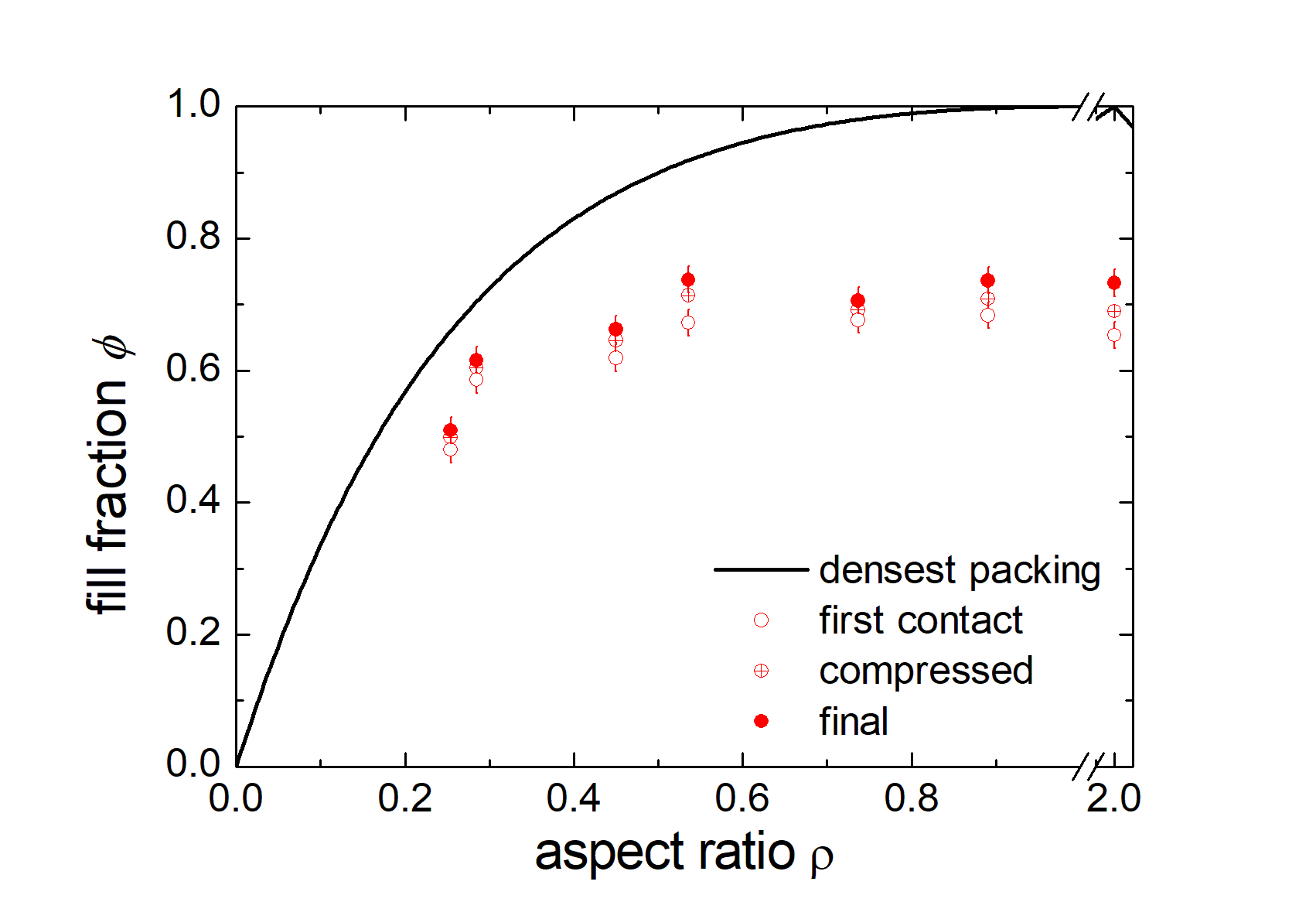}}

\caption{\label{fig:2}
a) Area filling fractions $\phi$ of several regular (crystalline) packings of crosses with different aspect ratios $\rho$. The black curve shows the densest packed
structure found. The individual packing fractions exist in the following ranges: TSQ and SQ for all $\rho$, ETE for $\rho<2$, SBS for $\rho>2$, FSQ exists up to $\rho \approx 0.52$. b) Actual experimental fill fractions with randomly packed crosses (symbols see text).
}
\end{figure}

The densest packing structure for crosses with small aspect ratios $\rho=b/a$ is the tilted square, TSQ: Two arms of adjacent crosses lay next to each other side by side. The unit cell is indicated by a dashed square in the Figure. It is tilted respective to the arms of the cross by an angle $\beta = \arctan [b/(a+b)] =\arctan[\rho/(1+\rho)]$. There is one cross per unit cell, as in all other configurations described in this section.
The length of the vectors spanning the unit cell is $r_{\rm T} = \sqrt{(a+b)^2+b^2}$. Since the area of a single cross is $A_0=4ab+b^2 = a^2 (4\rho+\rho^2)$, and the area of the unit cell is $A_{\rm T}=(a+b)^2+b^2=a^2 [(1+\rho)^2+\rho^2]$. The packing fraction is
\begin{equation}
\phi_{\rm (TSQ)}= \frac{A_0}{A_{\rm T}} = \frac{4\rho+\rho^2}{(1+\rho)^2+\rho^2}  .
\end{equation}
It reaches its maximum value one for an aspect ratio $\rho=1$ (see Fig.~\ref{fig:2}). When $\rho$ increases further, it decays again.
The graph of $\phi_{\rm (TSQ)}$ is shown in Fig.~\ref{fig:2}a as a black solid curve in the range $\rho<(1+\sqrt{5})/2$ where it provides the best area filling of all configurations considered here, and as a blue dashed line for larger $\rho$ where other arrangements are superior. It reaches the asymptotic value 1/2 for $\rho\rightarrow \infty$.

Another candidate for high packing fractions is the rhombic unit cell shown in Fig.~\ref{fig:1} for $\rho=1.5$. We label it ETE (end to end), since
two crosses touch each other with the ends of their arms, e.~g. the green and orange one in the figure, and two crosses sidewards, e.~g. the cyan and yellow ones, touch the arms of these
crosses from the sides. The unit cell area is equal to that of the rhomb drawn in Fig.~\ref{fig:1} for $\rho=1.5$, with side lengths $a+b$
and $2a+b$, the area is $A_{\rm E} = a^2 [(1+\rho)^2 + 1+\rho]$, and the corresponding fill fraction
\begin{equation}
\phi_{\rm (ETE)}= \frac{A_0}{A_{\rm E}} = \frac{4\rho+\rho^2}{(1+\rho)^2 + 1+\rho}.
\end{equation}
For low $\rho$, this structure has a lower fill fraction than the tilted square configuration described above, but at fill fractions
$\rho > (1+\sqrt{5})/2\approx 1.618$ the packing is more efficient than all competing configurations. This ETE structure exists up to $\rho=2$ where
again, the perfect tiling $\phi=1.0$ is achieved. For $\rho>2$, another structure with a rhombic unit cell is found which is more efficient than
the tilted square TSQ. Crosses are stacked with arms side by side (SBS) alternating with their upper and lower arms, as seen in the $\rho=2.5$ example in Fig.~\ref{fig:1}.
The unit cell is again a rhombus, as shown in the figure, but one can also choose a parallelogram with a side length of $2b$ and a height $a+b$. The unit
cell area is $A_{\rm S}= 2(\rho+\rho^2)a^2$. The packing fraction becomes
\begin{equation}
\phi_{\rm (SBS)}= \frac{A_0}{A_{\rm S}} = \frac{4\rho+\rho^2}{2\rho+2\rho^2}.
\end{equation}
The ETE structure has one peculiarity that distinguishes it from the previous lattice. It consists of loosely coupled sublattices
and it has a soft mode: The two sublattices of crosses stacked end-to-end in a line can be shifted with respect to each other by displacements
$\pm (a-b/2)= \pm (1-\rho/2)a$ whereas the global packing fraction remains unchanged.

Finally, for $\rho>1+\sqrt{3}\approx 2.732$, another structure provides the optimum packing, which we will refer to
as tilted rhomboid, TRH. The vectors spanning the unit cell have lengths of
$\sqrt{(2a+b)^2+4a^2}$ and $\sqrt{(a+b)^2+b^2}$,
The packing fraction is
 \begin{equation}
\phi_{\rm (TRH)}= \frac{A_0}{A_{\rm TRH}} = \frac{4\rho+\rho^2}{\rho^2+4\rho+2}.
\end{equation}

Finally, we mention a candidate that does not reach the optimum packing fraction but is nevertheless interesting
in non-perfect configurations. It can be described by an non-tilted square unit cell, and we denote it by SQ.
The side length
of the unit cell is $r_{\rm SQ}=\sqrt{(2+\rho)^2 a^2}$, the area is $A_{\rm Sq}=(2+\rho)^2 a^2$. The structure is exemplarily shown for $\rho=4$
in Fig.~\ref{fig:1}. It has a packing fraction of
\begin{equation}
\phi_{\rm (SQ)}= \frac{A_0}{A_{\rm SQ}} = \frac{4\rho+\rho^2}{(2+\rho)^2}.
\end{equation}
(green dashes in Fig. \ref{fig:2}).
This structure is interesting because, unlike TRH, this one is infinitely degenerate. One can shift either individual rows or individual columns arbitrarily, keeping $\phi$ unchanged. This structure can be considered a smectic arrangement, with layers that can be shifted respective to their neighbors. For the optimum space filling, this structure is less suited than TRH at all aspect ratios, but if one considers systems with thermal motion, it may be entropically advantageous.

At $\rho\rightarrow \infty$ ($a=0$), the crosses degenerate to squares, and the packing fraction reaches $\phi=1$ for the third time.
The complete graph of the fill fractions corresponding to the regular structures defined above is shown in Fig.~\ref{fig:2}a.
The dashed lines show the fill fractions of the different configurations where they exist but are no optimum structures. The solid black line is the plot of the optimum configurations for given $\rho$.

Several other regular configurations with more than one cross per unit cell are shown in Appendix A. None of them reaches a higher packing fraction
than the optimum configurations described above.

We emphasize that we have only considered regular packing configurations. Even though there is great confidence that the most efficient packing structures have been identified \cite{Atkinson2012}, we have no mathematical proof that denser configurations cannot exist. The main idea of introducing these lower limits of densest packing here was to set a scale for the evaluation of the experimental results on randomly packed crosses.
This is described in the following sections.

\section{Random packings}
\subsection{Samples and experimental setup}

We use plastic crosses that are commercially available as spacers for tiling. They can be obtained in different aspect ratios. In the present study,
we selected crosses for nominal 2 mm, 3 mm, 4 mm, 5 mm, 8 mm and 10 mm arm widths, corresponding to aspect ratios between 0.25 and 0.89. In addition, crosses with
an aspect ratio of 2 were 3D printed in the lab. We refer to the latter shortly as black crosses.
Table \ref{tab1} lists the geometrical dimensions of these five species. 

\begin{table}[htbp]
\centering
{
		\renewcommand{\arraystretch}{1.5}
		\begin{tabular}  {|l |c c c c c |c|}
			
			\hline\hline
			 nominal &$\rho$      & $a$ [mm]   & $b$ [mm]      &   $\ell $ [mm]       &  $r_0$ [mm]  & rounded \\
			\hline
			3 mm  &\,0.248  &  12.1  & 3.0   &  27.2  &15.4 &no\\
			
			2 mm   &0.284  &  7.4  & 2.1   &  16.9  & 9.73&yes\\
			
			4 mm  &0.445 &  9.2    & 4.1   &  22.5  &13.9&yes\\

			5 mm  &0.516 &  9.5    & 4.9   &  23.9  &15.2&yes\\
			
			8 mm  &0.728 &  10.5   & 7.65  &  28.6   &  19.7&no\\
			
            10 mm &0.894 &  10.8   & 9.65  &  31.2   &  22.6&no\\

            black & 2.0 & 5.0 & 10.0 & 20.0 & 18.0&no\\
        \hline
  		\end{tabular}
	}

	\caption{Geometrical parameters for the plastic crosses used in our experiments.
The lengths are accurate within $\pm 0.1$~mm, thus we round the aspect ratios to two digits in the following plots.
The 2 mm, 4 mm and 5 mm crosses have arms with semicircular ends, the others are flat-ended. The arm length $a$ of the rounded crosses
includes the rounded end.
}
	\label{tab1}%
\end{table}

The height of the crosses is between 3 mm for the smallest type and 6 mm for the largest.
A static friction coefficient $\mu=0.3\pm 0.05$ of the crosses has been estimated by measuring the tilt angle of a plane on which a cross
starts to slide down from the surface of a second cross.

\begin{figure}[htbp]
\center
\includegraphics[width=0.95\columnwidth]{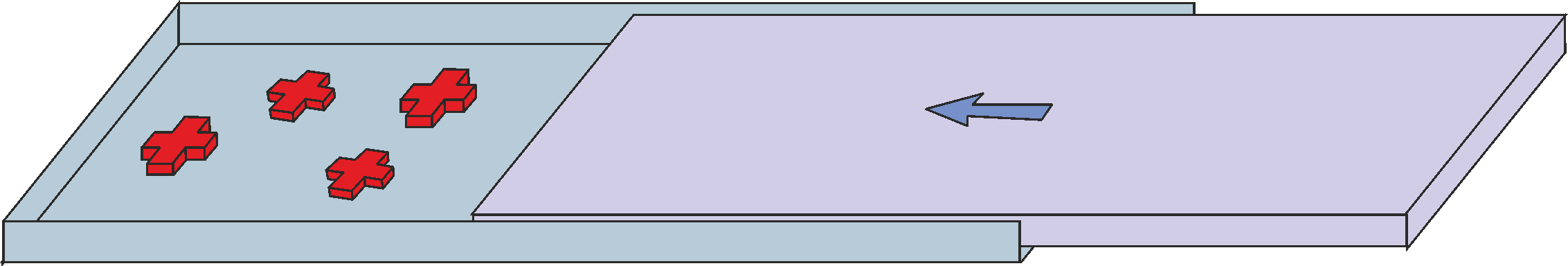}
\caption{\label{fig:setup}
Schematic drawing of the experimental setup with the table and slider.
}
\end{figure}

The crosses are initially placed at random on a horizontal plate with three fixed rails that form three sides
of a rectangular base (Fig.~\ref{fig:setup}). We leave enough space between the crosses so that
the initial area filling fraction is at least 2 to 3 times larger than in the jammed state, and we try to prepare a
rather uniform distribution. Each experiment starts with a new initial configuration.
After placing the crosses, this area is covered by a transparent glass plate. At the fourth side of the rectangular area, a slider can be manually moved in to
shift the crosses and to compact the ensemble. The slider is first moved to a position where the crosses resist the advancement and a
jammed state is reached. Then, a
photo is taken. The slider is then pushed further with a force of a few Newton, which further leads to a slight compaction of
the crosses configuration. Thereafter, a second photo is taken and the slider is then hit hard several times, in order to cause further
compaction. Then, we reach a state where further hitting the moveable side bar does not lead to an improvement of the fill fraction and
a third photo is taken. The jamming point is assumed to be in the range between the first and third slider positions.
Strictly, the first point where the slider experiences resistance should be defined as the jamming point of our system, but since this
first contact is often established by only one or very few crosses, such a definition would slightly underestimate the
behavior of an infinitely large system.
Image analysis is used to determine the positions and orientations of all crosses. From these data, we can not only determine the packing
fractions, but also the spatial correlations between cross centers and their mutual orientations.
The experiments were performed with sets containing between 166 and 500 crosses. The number of crosses and the number
of repetitions of each experiment are listed in Table~\ref{tab:2} in Appendix C. \\

\begin{figure}[htbp]

\includegraphics[width=0.49\columnwidth]{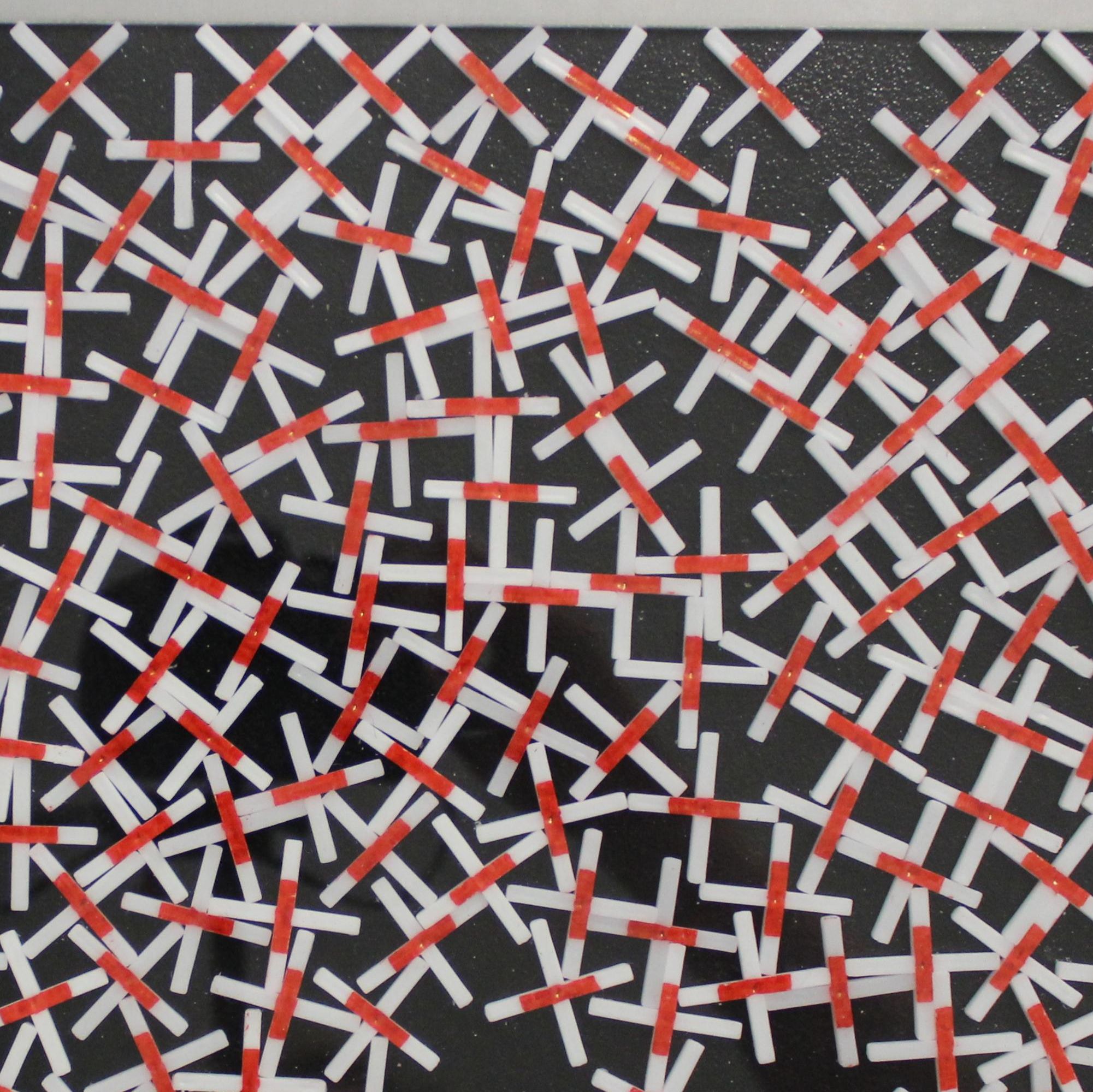}
\includegraphics[width=0.49\columnwidth]{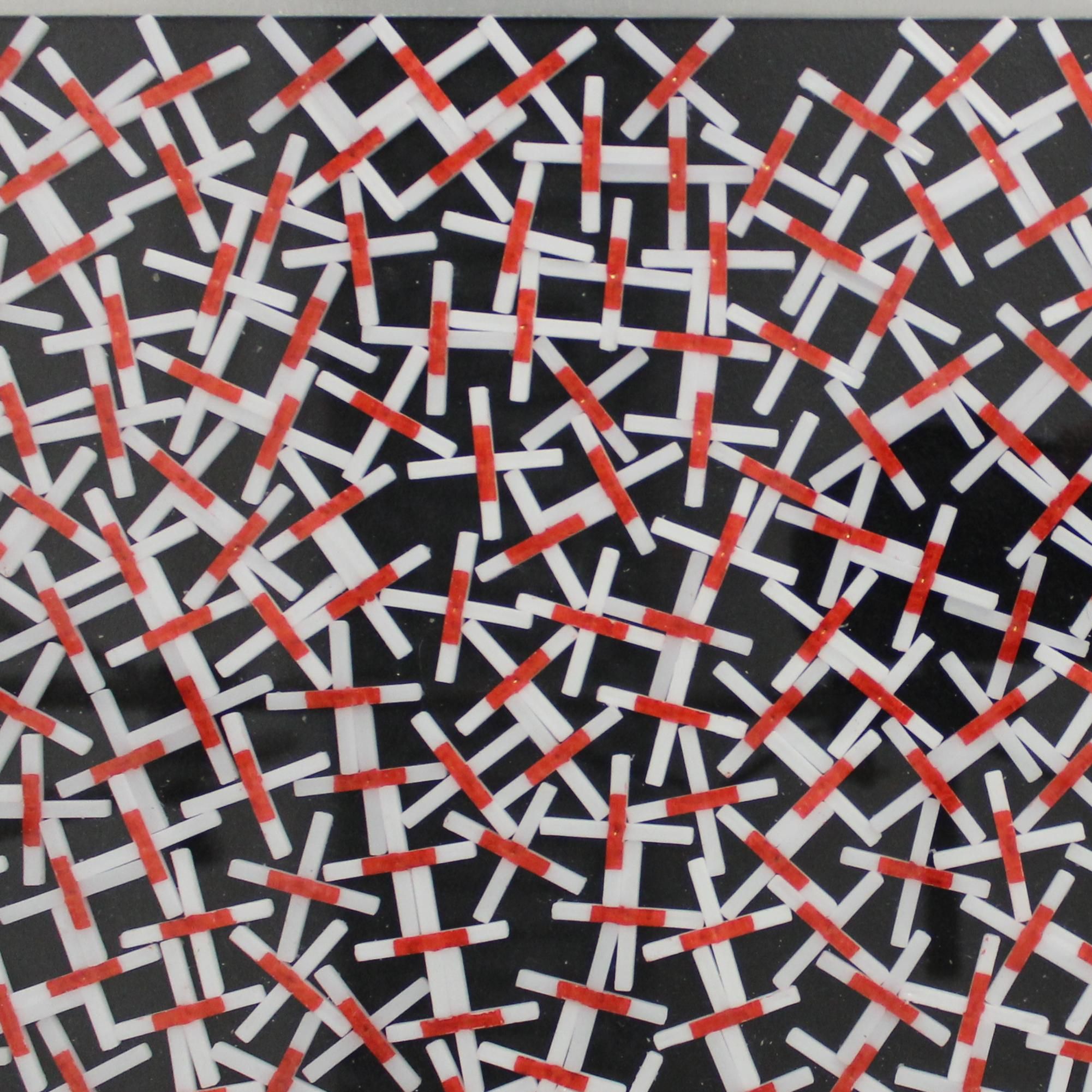}

\includegraphics[width=0.49\columnwidth]{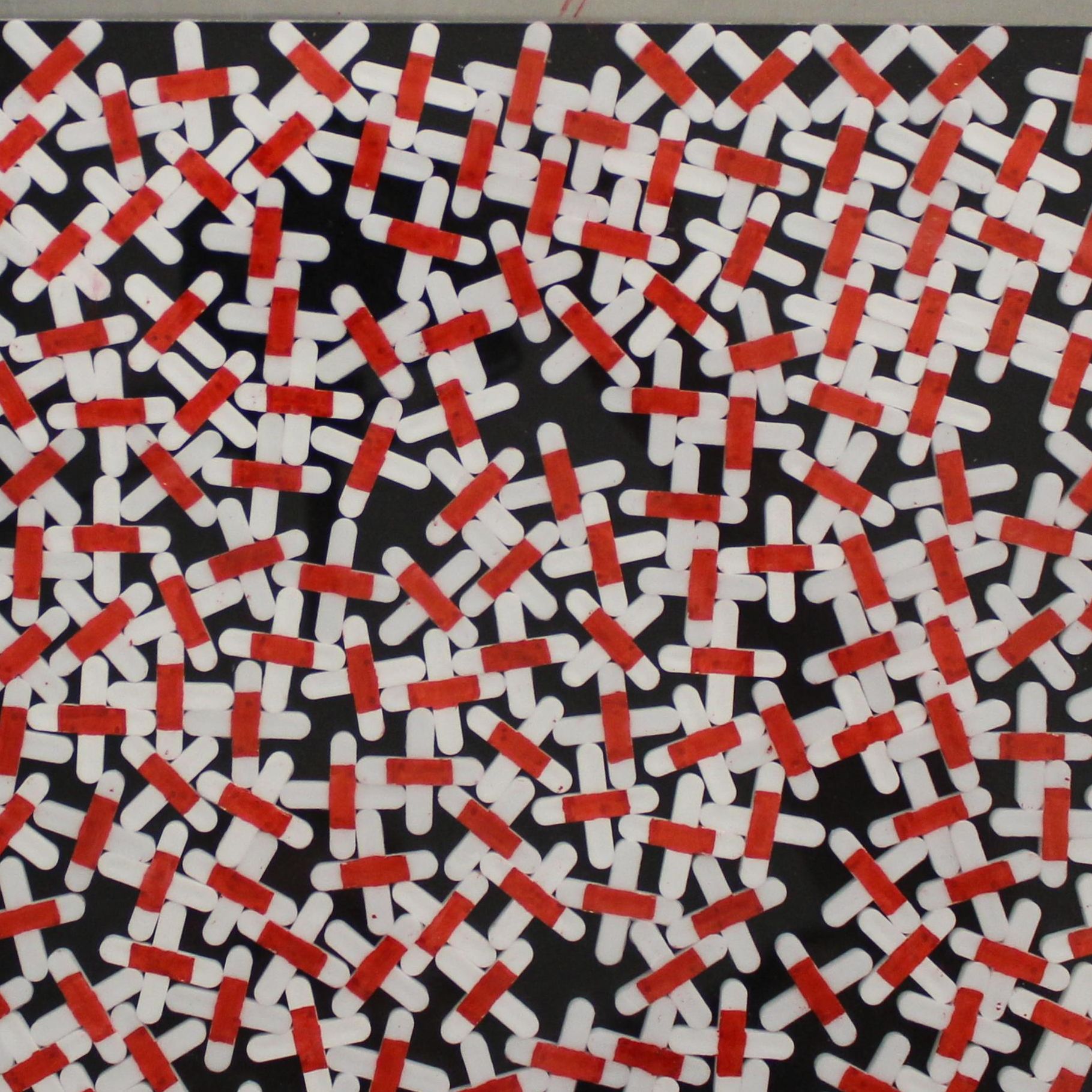}
\includegraphics[width=0.49\columnwidth]{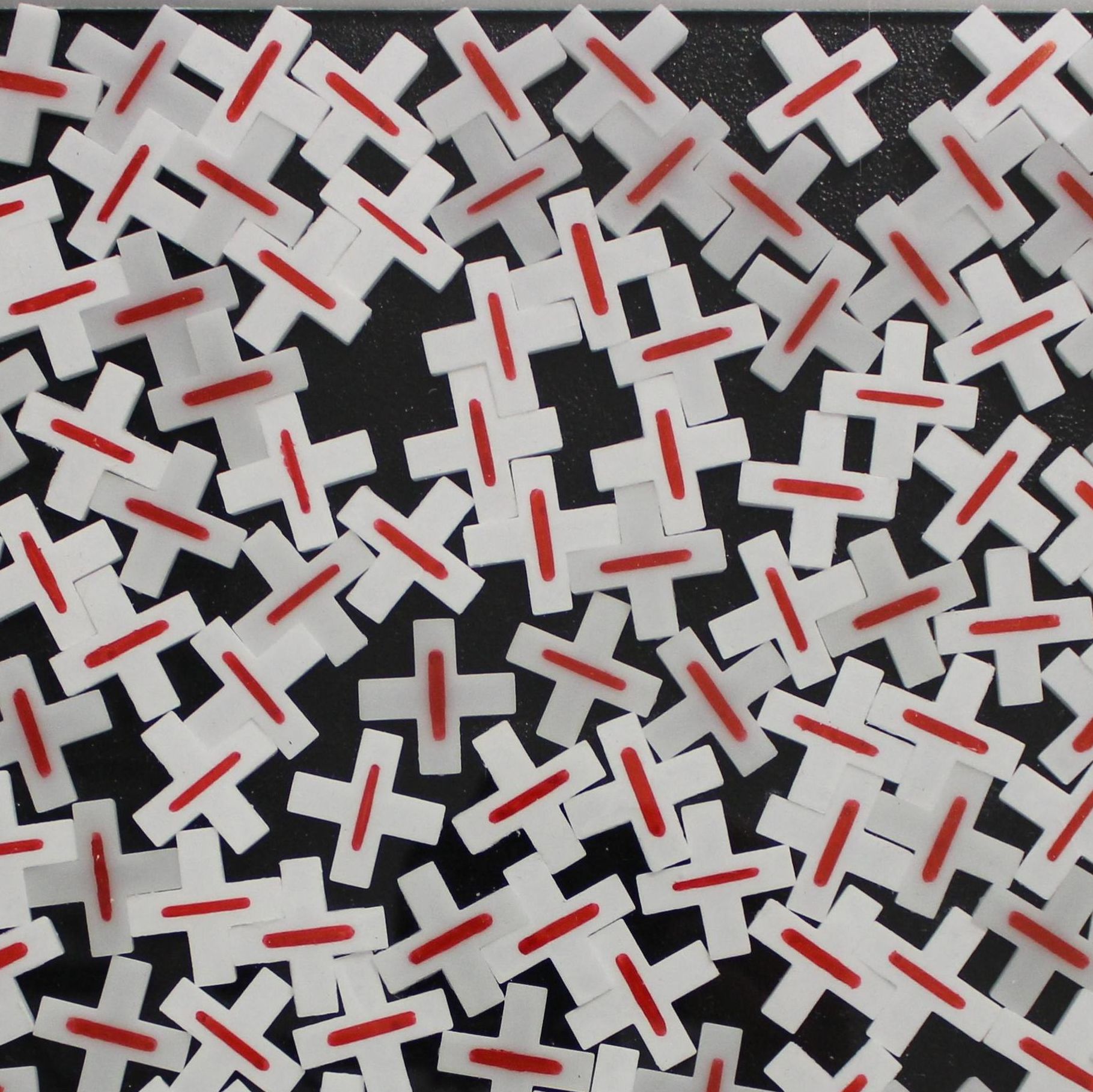}
\caption{\label{fig:images}
Images of randomly packed crosses. Top left: 3 mm crosses after first phase, loosely packed; Top right: 3 mm crosses after final compaction;
Bottom left: jammed 5 mm crosses (arms with rounded ends); Bottom right: jammed 8 mm crosses. All images show a small 20 cm $\times$
20 cm clip next to the moving edge (top).
Red labels merely serve as identifiers to facilitate image processing.
}
\end{figure}

\begin{figure}[htbp]

\includegraphics[width=0.48\columnwidth]{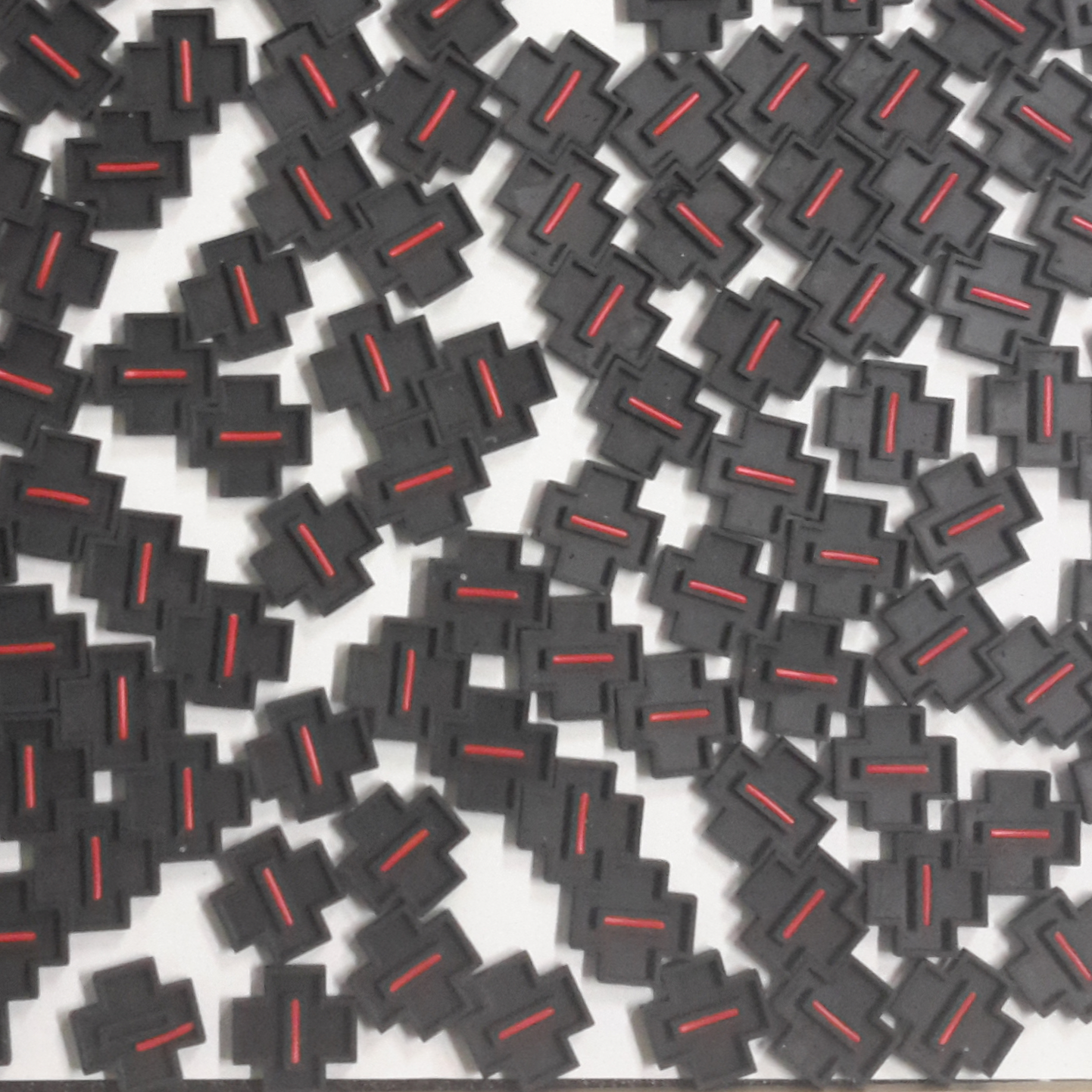}
\caption{\label{fig:black}
Image of randomly packed crosses with aspect ratio $\rho=2$. The positions and orientations of the crosses are identified
by means of the red marker bars.
}
\end{figure}

\subsection{Area filling fraction}

The crosses are placed on a plane in a loose packing and then compressed laterally by a moving boundary as described above. Thereby, the
optimum space filling structure is never reached, even when the system is repeatedly shaken.
This is comparable to the problems of 3D packing of objects, where an optimum space filling is usually failed by far.
Small clips of typical photos of the experiment are shown in Figs. \ref{fig:images} and \ref{fig:black}. 
In Fig. \ref{fig:images}, one notices that the crosses in contact with the
moving slider are preferentially aligned diagonal to the edge, touching it with two arms, but this induced order does not penetrate
more than one to two cross diameters into the plane. This issue is discussed in more detail later. The two top images show the same area
in two configurations.
The left one shows the crosses after they first block the slider motion. The right one presents the same experiment after the slider was
pushed hard several times towards the crosses, until no further rearrangement could be noticed. There is surprisingly little difference
between the statistical properties of both configurations, a feature that is discussed next:

\begin{figure}[htbp]
\centerline{
\includegraphics[width=0.85\columnwidth]{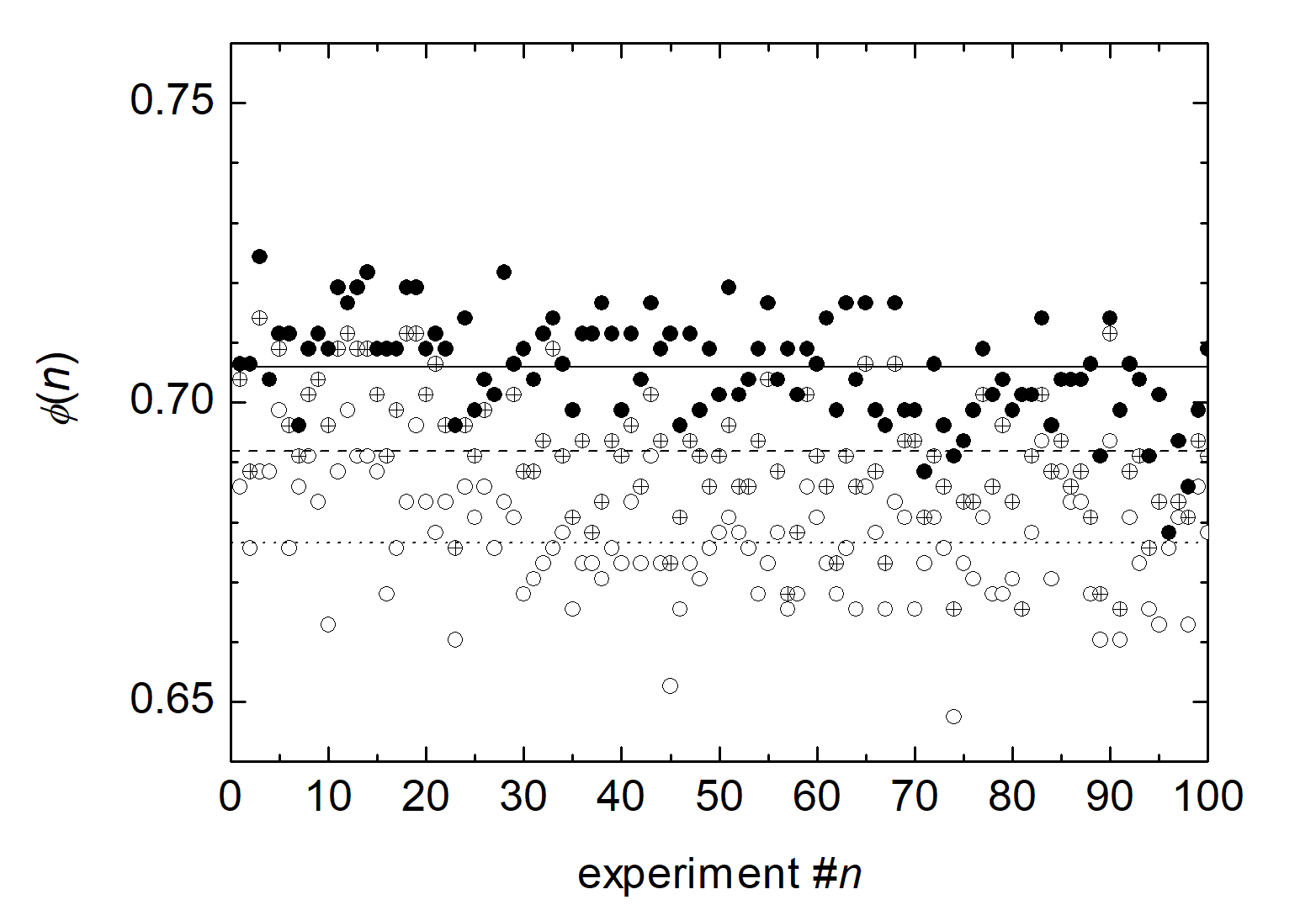}}
\caption{\label{fig:4}
The plot shows the packing fractions $\phi$ determined in 100 independent experiments (crosses with $\rho = 0.73$). The symbols correspond to the value measured when the first
blockage of the movable bar is reached ($\circ$), the value when the sample is further compactified with larger force ($\oplus$),
and the compaction reached after hard pushes of the bar ($\bullet$). The corresponding averages are plotted as dotted, dashed and solid lines, respectively.
}
\end{figure}

Figure \ref{fig:4} shows the result of 100 independent experiments (300 images) with 8 mm crosses ($\rho=0.73$). We present three graphs, where
the open symbols show the packing fraction $\phi$ after the first compaction phase, the filled symbols the packing fraction after
the system is repeatedly hit from the side until no rearrangements occur any more. The packing fraction or area fill fraction $\phi$ is calculated
by the area covered by crosses divided by the total area. The averages of these measurements differ by less than 5\%.
Figure \ref{fig:2}b compares the packing fractions of random close packing (jamming point) of crosses of different aspect ratios with the curve
of the optimum regular packing, TSQ for $\rho<1$. Two trends are evident in the figure: First, the random packing fractions are much smaller
than the regular optimum packing fractions for given aspect ratios (roughly 70\% of the latter). The fill fraction $\phi(\rho)$ increases with
increasing aspect ratio. The second trend is that crosses with rounded ends (2 mm, 4 mm and 5 mm crosses in our study) apparently
achieve a somewhat larger fill fraction than one would expect by interpolation of the data of flat-ended crosses.
This is intuitively clear: When
crosses with flat ends touch each other with edges, further rotations are blocked unless the distance between the cross centers
is increased again. With semicircular ends, the crosses can rotate with respect to each other to find denser configurations.

\subsection{Spatial correlations}

\begin{figure}[htbp]
\centerline{
  \includegraphics[width= \columnwidth]{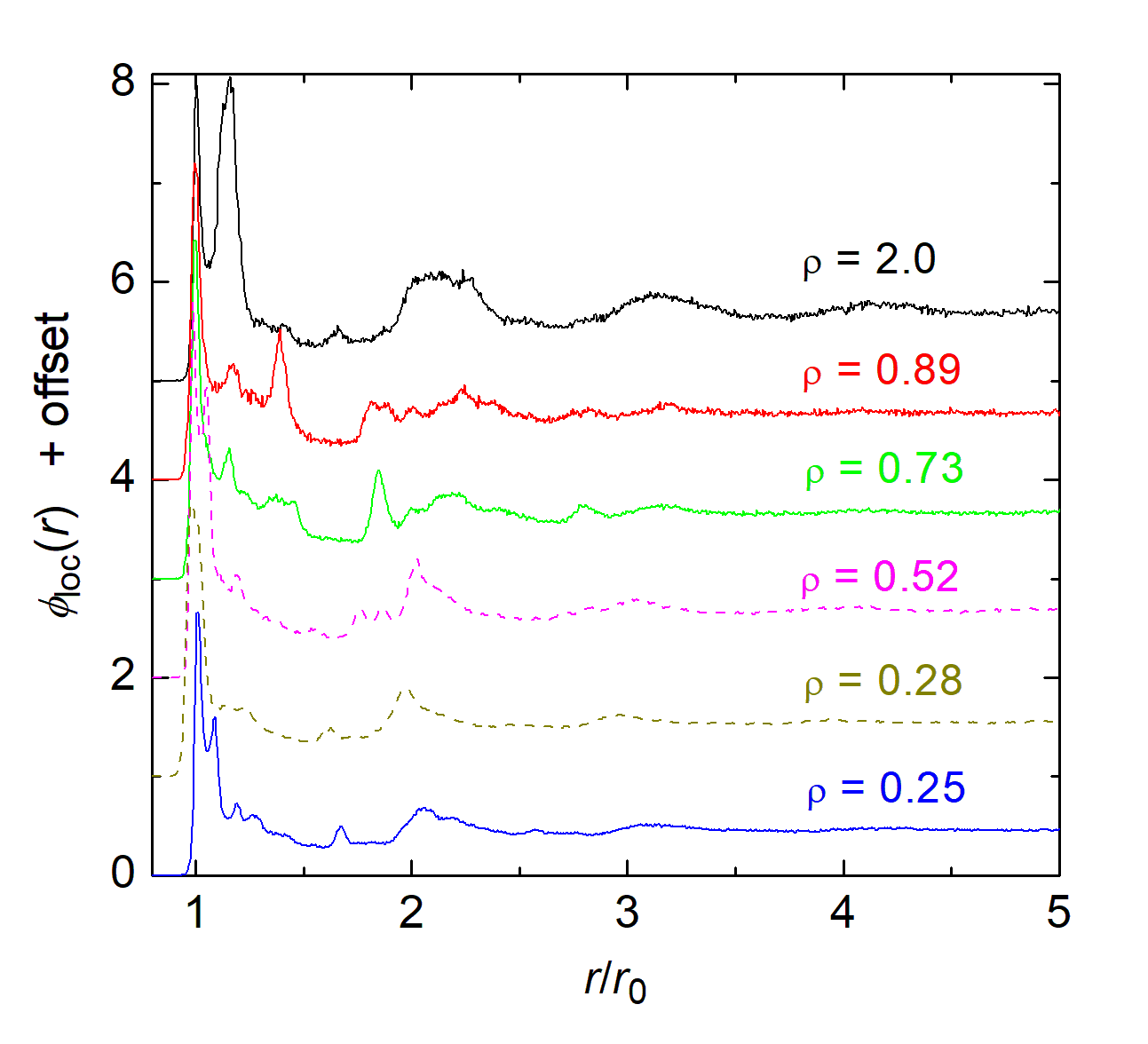}}
\caption{\label{fig:3}
Experimental results for the correlation of densities
in dependence on the aspect ratio. The effective local fill fraction $\phi_{\rm loc}$ is the product
of the average number $N(r)$ of cross centers per area in a distance $r$
from a central cross, multiplied with the cross area $A_0$, and $r_0$ is the minimal possible distance of neighbors.
In order to separate the graphs, we added offsets of 0, 1, 2, \dots 5 from bottom to top.
Selected individual graphs are shown in Figs.~\ref{fig:commented} and \ref{fig:blackcorr} together with a peak assignment.
}
\end{figure}

Irrespective of the overall random arrangement of the crosses, one finds a strong correlation of orientations and positions
of neighbors. This is a consequence of the non-convex character of the particles. Figure \ref{fig:3}
shows the spatial correlation functions $\phi_{\rm loc}(r)$, which are proportional to the probabilities of finding crosses in a distance $r$ (center-center)
from a given cross. The spatial distances in these graphs have been normalized with the minimum
\begin{equation}
r_0 (\rho)= \sqrt{[(a+b)^2+b^2]}=\sqrt{[(1+\rho)^2+\rho^2]}\cdot a,
\end{equation}
defining the center-center distance of two crosses laying with their arms side by side (like those at the sides of the TSQ unit cells, Fig.~\ref{fig:1}). For a clearer presentation,
we have multiplied the number densities $N(r)$ of crosses with the area $A_0$ of the single cross to obtain $\phi_{\rm loc}$. Thus for large $r$,
where the cross positions are uncorrelated, the mean area fill fractions $\phi$ shown in Fig.~\ref{fig:2}b are reached asymptotically,
$ \lim\limits_{r\to\infty} \phi_{\rm loc}(r) = \phi$.

\begin{figure}[htbp]
\center

a) \includegraphics[width=0.9\columnwidth]{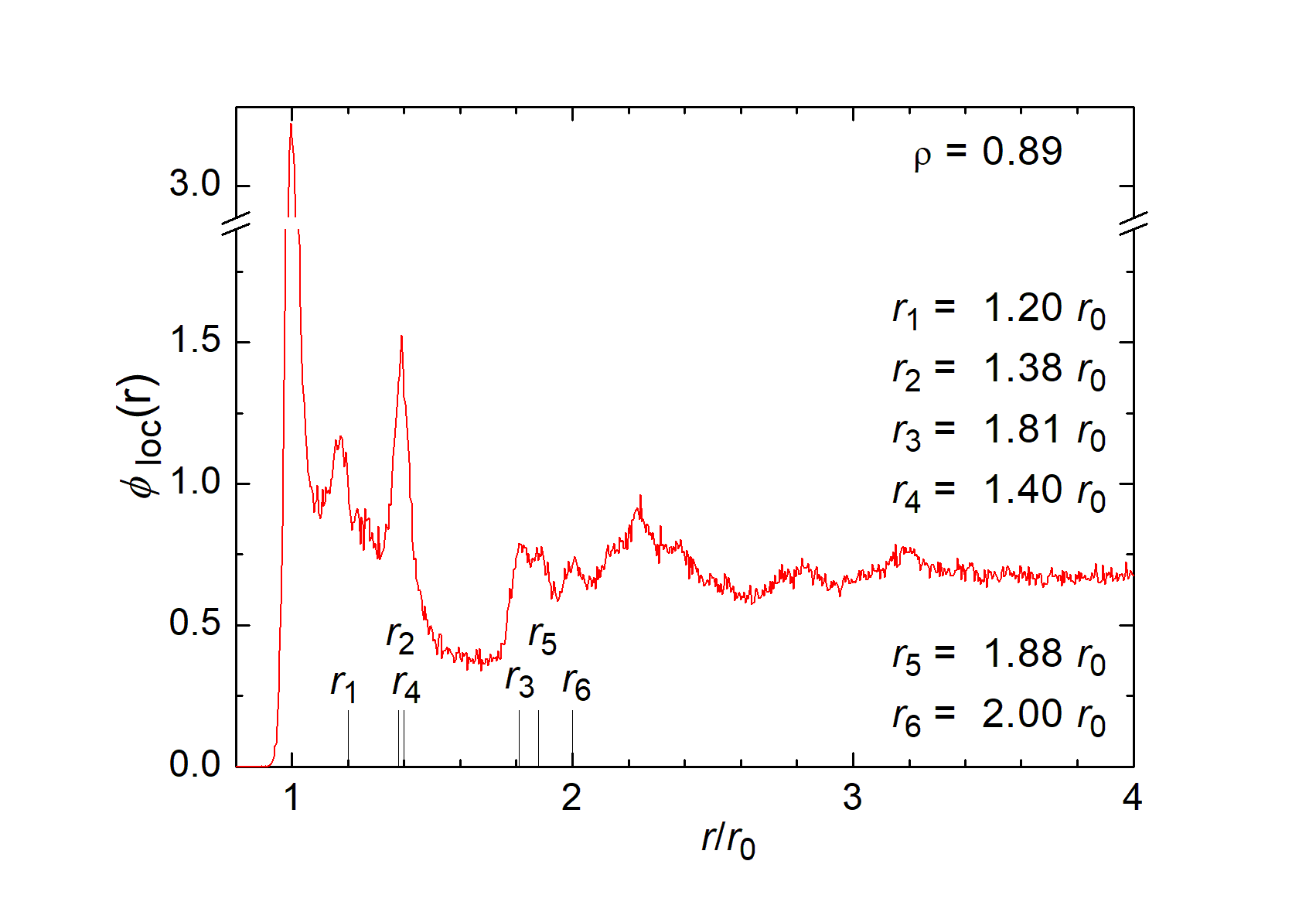}\\
b) \includegraphics[width=0.9\columnwidth]{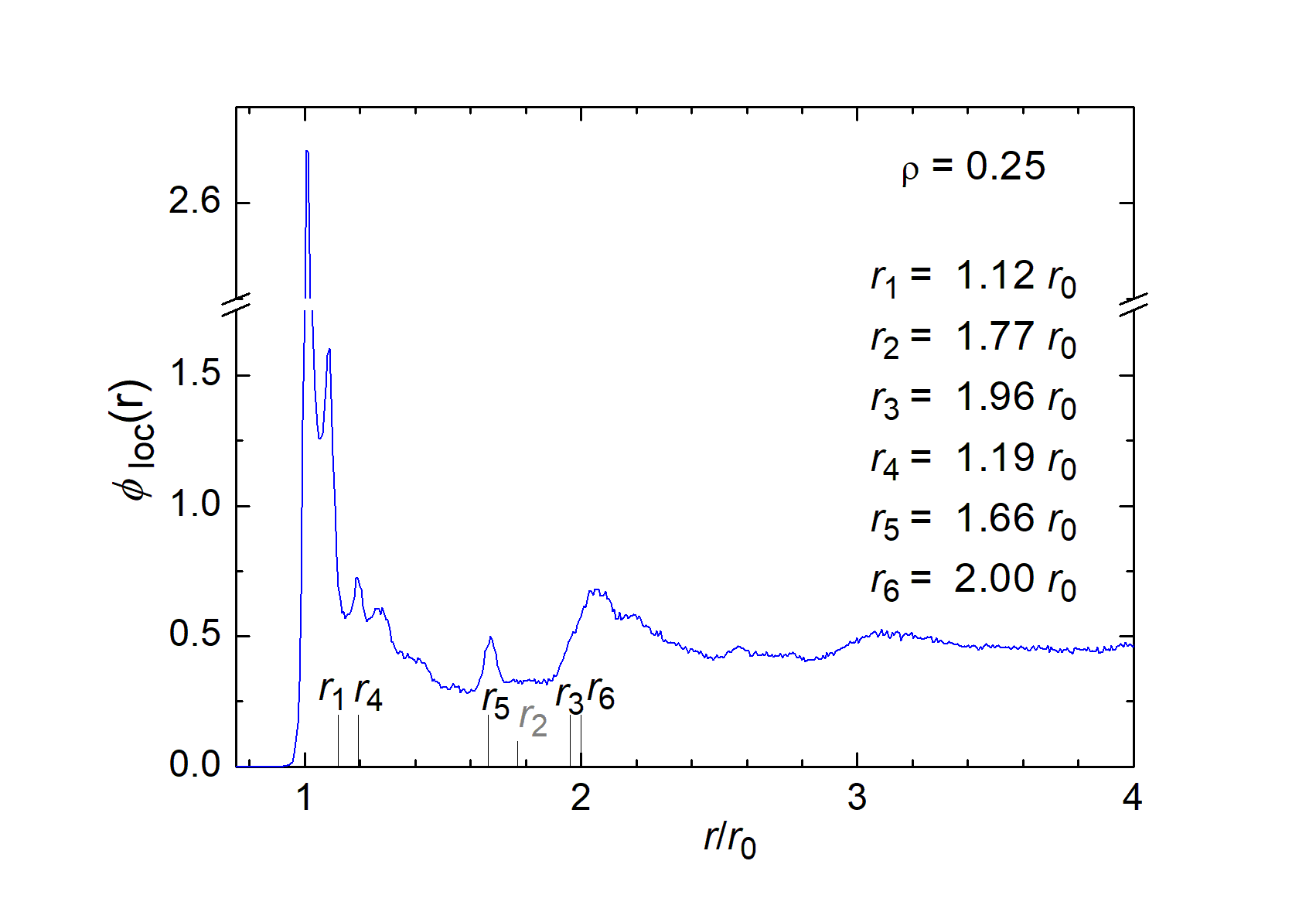}
\centerline{
c) \includegraphics[width=0.38\columnwidth]{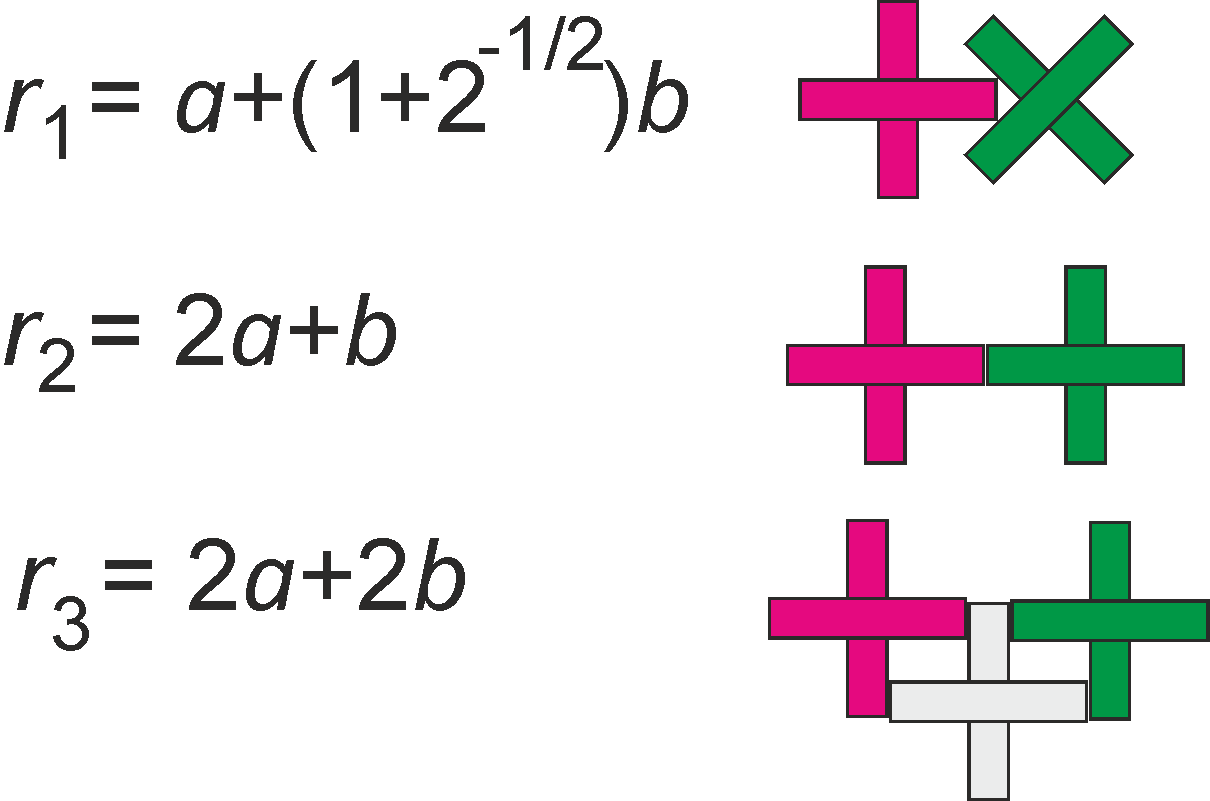}\hspace{5mm}
  \includegraphics[width=0.44\columnwidth]{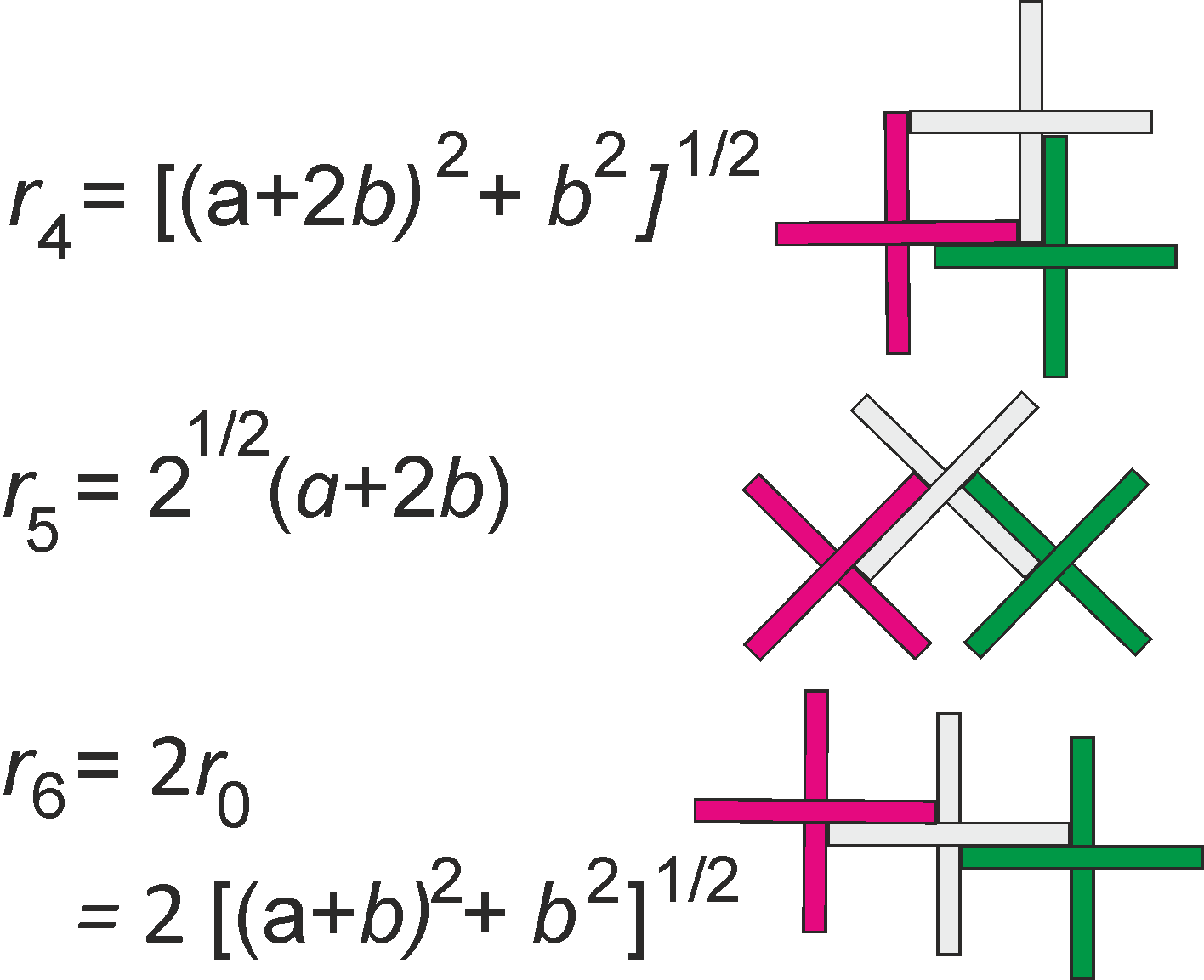}}
\caption{\label{fig:commented}
Graphs a) and b) are identical with those in Fig.~\ref{fig:3}.  c) shows the mutual positions and orientations of
pairs that contribute dominantly to the corresponding peaks
}
\end{figure}

In addition to the dominant maxima near $r= r_0$ in each graph, one identifies several more peaks in the individual graphs.
Their positions depend on the aspect ratios. Some of them are found in the vicinity of $2r_0$, and all
$\phi_{\rm loc}(r)$ graphs have a pronounced dip in the region around $1.5 r_0 <r<2.0 r_0$, these distances are clearly
under-represented. The cross positions show noticeable correlations over distances of about 3 to 4 $r_0$, i.~e. about two cross diameters, particularly for the crosses with broader arms (larger $\rho$).
Since there is no obvious relation between the aspect ratios $\rho$ and the relative peak positions at first glance,
their explanation needs a deeper analysis.
We solved this problem at least partly, by identifying pairs of crosses in the images that are separated
by distances near the peak positions. This reveals that some of the peaks, viz. $r_1$ to $r_6$ in Figs.~\ref{fig:commented}a and b,
are representative for typical pair configurations. Figure ~\ref{fig:commented}c sketches some of these configurations,
where the green and purple crosses are separated by $r_i$. With light grey we marked intermediate crosses that
stabilize the mutual orientations of the outer two.

Peaks near these six $r_i$ are primarily caused by a preference of the sketched local arrangements.
The peak $r_2$ caused by end-to-end contacts is practically absent in the case of thin crosses, as seen
in Fig.~\ref{fig:commented}b.
Note that all ratios $r_i/r_0$ depend on $\rho$. Therefore,
peaks of similarly correlated pairs have different relative positions.

\begin{figure}[htbp]
\center

 \includegraphics[width=0.75\columnwidth]{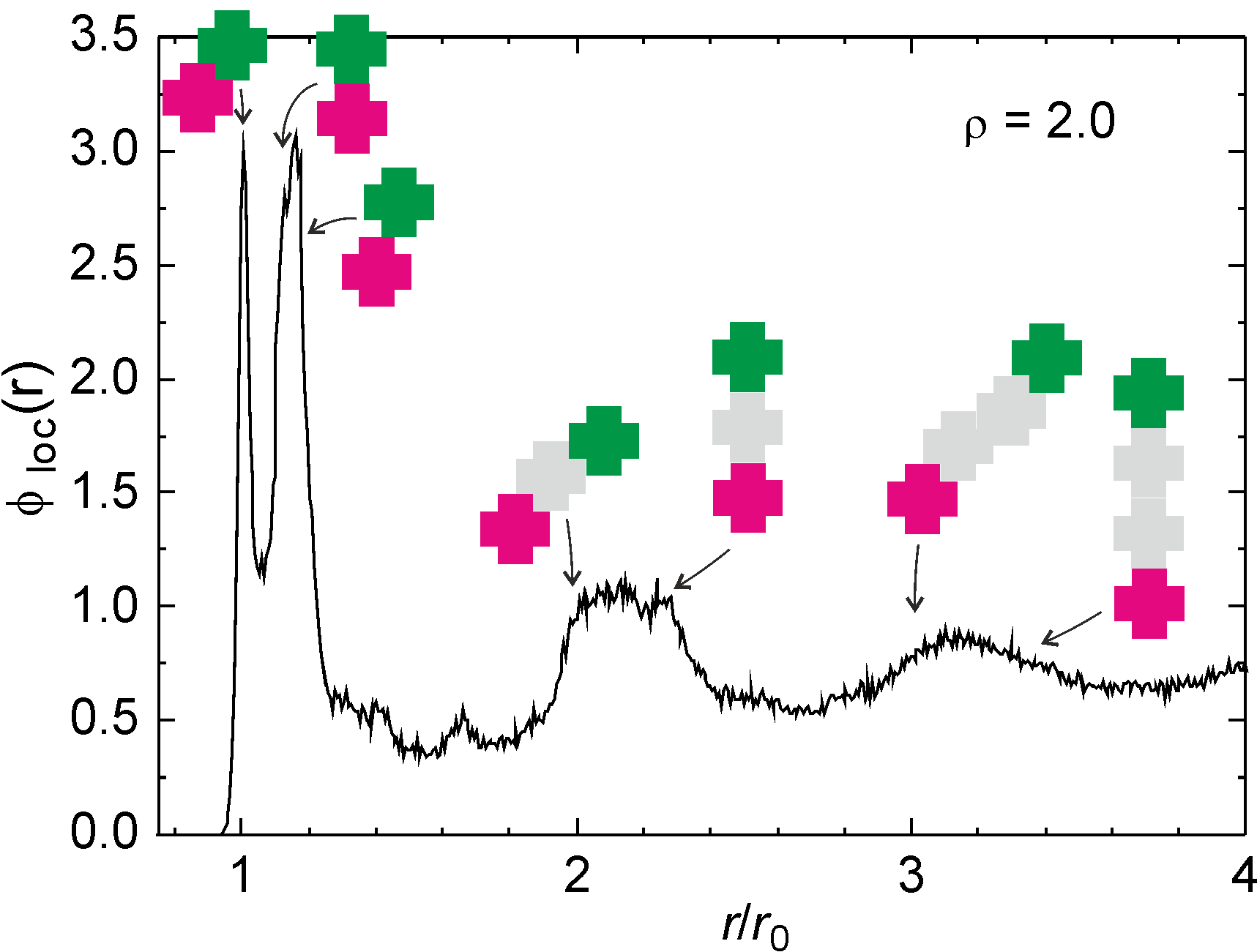}

\caption{\label{fig:blackcorr}
Graphs of the spatial correlation of cross centers for the aspect ratio $\rho = 2$. Note the pronounced differences to the
previous samples. The minimum distance is $r_0 = 18$~mm. The large peak starting near $20$~mm $\approx 1.1~r_0$ is generated by crosses
stacked end-to end. The configurations mainly responsible for the peaks are indicated by inset images.
Note that there are well-pronounced peaks of the second and third coordination spheres.
}
\end{figure}

Figure \ref{fig:blackcorr} shows the spatial correlations of crosses with $\rho=2$ (3D-printed black crosses). Two essential new features are seen:
First, a very pronounced peak appears next to the first maximum $r=r_0$. It is identified as the separation of crosses in end-to-end configuration, $r_2$.
When they are perfectly aligned, the peak appears at $\approx 1.1~r_0$ for $\rho=2$. When the crosses are shifted sidewards respective to each other,
the distance is larger, therefore this peak is rather broad. The second obvious feature is the pronounced appearance of peaks of the second and third coordination shells which are much less pronounced for crosses with $\rho <1$.

\subsection{Orientational correlations}

When discussing orientational order, the first question that needs to be clarified is whether the uniaxial compaction method, pushing one side bar,
has an influence on the global ordering of the crosses. Is the configuration statistically isotropic, or is there an induced uniaxial order?
Zheng et al.~\cite{Zheng2017} have compared the uniaxial compression method with the results obtained by biaxial compression, and they have found slight differences between the two jammed states. The jamming point is slightly shifted to larger packing
fractions for biaxial compression (see above), but during cyclic compression, the results of the two compression methods seem to converge. The biaxial compression is referred to as 'isotropic' in that paper, indicating that the authors assume a
configuration where all spatial directions are equivalent. Yet it was not explicitly studied whether uniaxial compression has an impact on the isotropic ordering of the ensemble or whether it creates a 'memory' of the compression history with a preferential axis.

Our Fig.~\ref{fig:images} indicates that at least at the borders, surface induced alignment is present, but we tried to find
a quantitative measure for the anisotropy of the orientation of crosses in the central part. For such a quantitative description, we use the global tetratic bulk order parameter
\begin{equation}
q_4 = \langle \cos 4\alpha_i \rangle
\label{eq:q4}
\end{equation}
where the $\alpha_i$ are the angles of the individual crosses respective to the sides of the rectangular box.
The coordinate system is chosen such that this angle is zero when the arms are along or perpendicular to the sides formed by the confining rails.
When computing $q_4$, we average over all crosses except those in the first layer near the rails. More accurately, we exclude all crosses whose
centers are closer than one cross size to any of the side walls. We further average over all images for a given cross type.
The result is that independent of the cross geometries, and independent of the state of compaction, all global order parameters $q_4$ of thin
crosses ($\rho < 1$) were very small but systematically negative. Those crosses are somewhat more likely to be diagonal to the borders than parallel to them.
The same is reflected in a prevalence of this orientation in the boundary layers (see, e.~g. Fig.~\ref{fig:images}, top). The averages of $q_4$ over
all experiments were of the order of $-0.01$ for all shapes of thin crosses. This means that the compaction method used here (uniaxial compression)
had negligible influence on the global orientational order. The configurations can be assumed to be isotropic in reasonable approximation.
The wide crosses with $\rho=2$ showed an equally weak orientational order, but with a positive sign of the global order parameter $q_4$. Here,
the average was $q_4=0.014$. This value is still tolerably small to consider the system fairly isotropic, but the sign change
shows that the crosses were somewhat more likely to be aligned with their arms parallel or perpendicular to the sliding bar rather than diagonal.
This slight prevalence is induced by our preparation method of the jammed states. A simple explanation of this mechanism
is given in Appendix B. From the arguments given there it is plausible to assume that similar effects will be found
under biaxial compression, because the order induced at the fixed and moving bars are comparable.

\begin{figure}[htbp]
\center
\includegraphics[width=0.95\columnwidth]{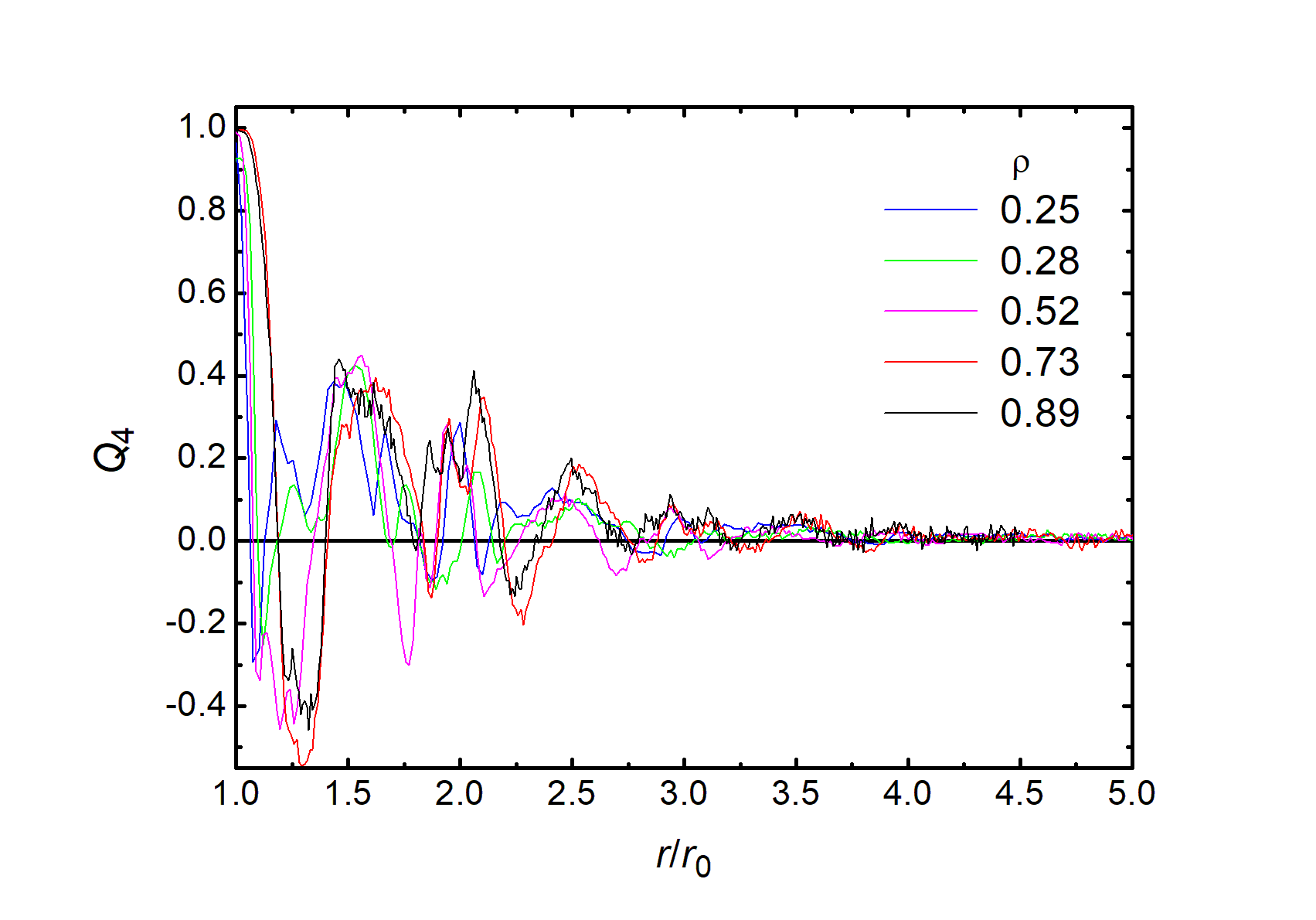}
\caption{\label{fig:order}
Local order parameter $Q_4(r)$ of the surrounding crosses in distance $r$ from a central cross, averaged over all central
crosses.
}
\end{figure}

\begin{figure}[htbp]
\center
 a) \includegraphics[width=0.85\columnwidth]{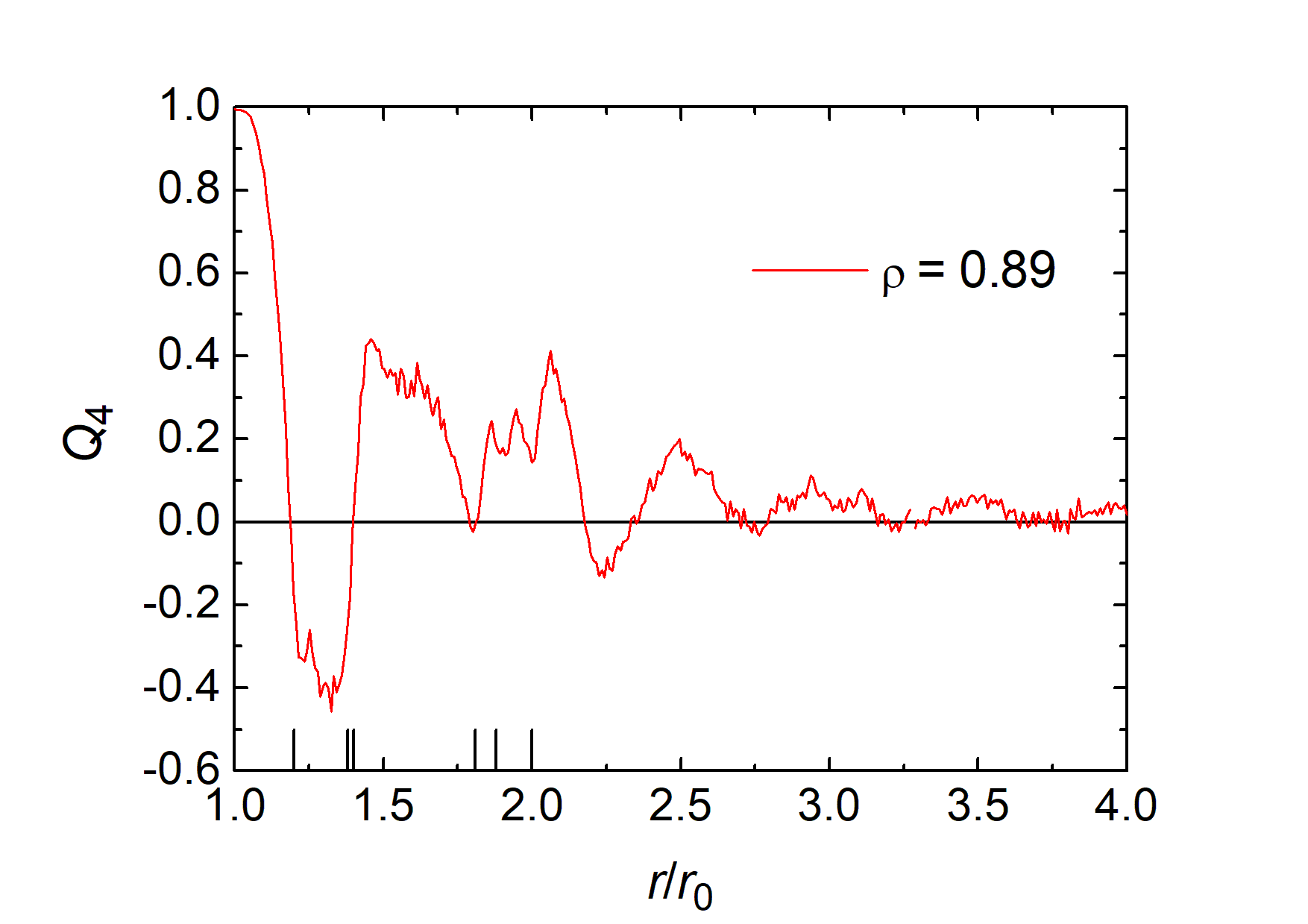}\\
 b) \includegraphics[width=0.85\columnwidth]{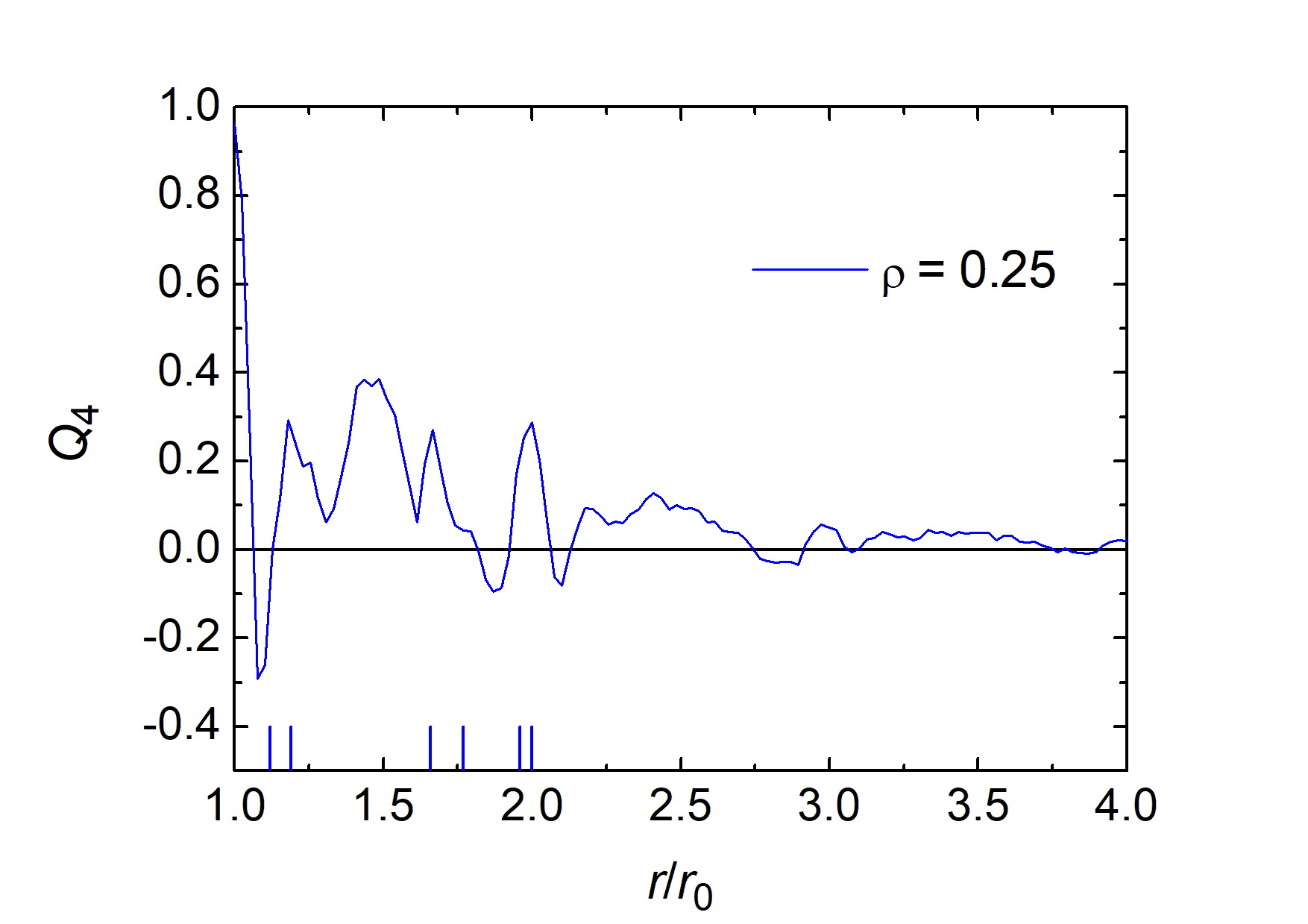}
 \caption{\label{fig:order2}
Local orientational correlations parameter $Q_4(r)$ of the surrounding crosses in distance $r$ from a central cross for
a) $\rho = 0.89$ and b) $\rho = 0.25$. Markers indicate the labeled peak positions of $\phi_{\rm loc}$ in
Figs.~\ref{fig:commented}a,b.
}
\end{figure}

Irrespective of the global isotropy, neighboring crosses show substantial correlations of their orientations.
This is reflected in the plots in Fig. \ref{fig:order}.
Here, we computed the local orientational correlation parameter
\begin{equation}
Q_4 (r) = \langle \cos 4\varphi_i \rangle .
\label{eq:Q4}
\end{equation}
Here, we average over all crosses in distance $r$ from the central cross and $\varphi_i$ are the angles of the crosses $i$ in
distance $r$ from the central cross with respect to the orientation $\alpha_0$ of the latter: $\varphi_i=\alpha_i-\alpha_0$.
Because of the fourfold symmetry of the crosses, these angles
are taken in a 90$^\circ$ range. This order parameter $Q_4$ reaches its maximum $+1$ when two crosses are oriented with their arms parallel,
and the minimum $-1$ indicates that the second cross is diagonal to the first one. With 150 to 300 experiments for each $\rho$ and between 150
and 250 crosses in each image, the statistics are based on data of at least $10^6$ pairs.

\begin{figure}[htbp]
\center
 \includegraphics[width=0.95\columnwidth]{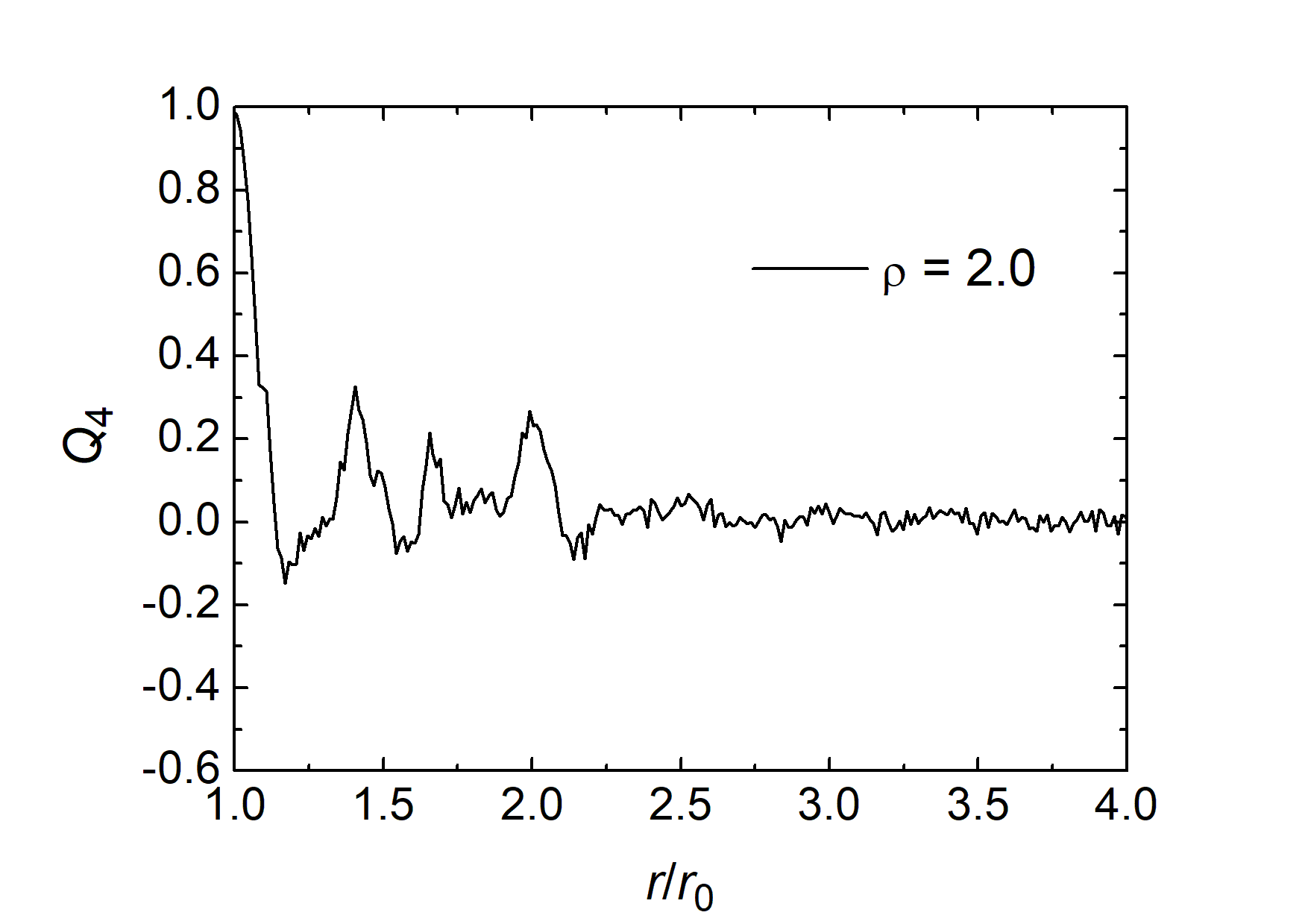}

 \caption{\label{fig:orderblk}
Local orientational correlations parameter $Q 4(r)$ of the crosses in distance $r$ from a central cross for $\rho=2$.
}
\end{figure}

It is evident that for distances close to $r_0$, only aligned crosses can exist. Thus $Q_4(r_0)=1$ for all types of crosses.
Then, the local order parameter drops to negative values because crosses in distances near $r_1$ preferentially
contribute to a negative value of the $Q_4$. This orientational order is less perfect than that of the directly attached crosses, thus
the negative peak reaches only values around $-0.5$ (for the thin crosses). The orientational correlation $Q_4(r)$ changes its sign with
increasing $r$ several times. It finally vanishes over a distance of about $4~r_0$,

The large undulations of $Q_4(r)$ with $r$ are systematic, they represent characteristic features of the random packing structures.
However, there appears to be no clear correlation between the peak positions in $\phi_{\rm loc}$ and those in the local orientational order, except the
global maximum $Q_4(r_0)\approx 1$ and the local minimum $Q_4(r_1)$.
Again, the wide crosses ($\rho=2$) shown in Fig.~\ref{fig:orderblk} are an exception. There, the negative peak is suppressed, and the parameter
$Q_4(r)$ is predominantly positive almost everywhere. The probability that the crosses in a certain distance are diagonal to the central cross
 is much lower than for the thin crosses.
A pronounced positive peak at $r=2r_0$ indicates aligned crosses in the 2nd coordination sphere of the central cross.

\section{Discussion and summary}
\label{sec:disc}
A few literature results obtained in numerical simulations can be used as benchmarks for our experimental results.
Concerning the densest crystalline packing, our predictions for $\rho < 1$ can be compared with results reported by Meng et al.~\cite{Meng2020}
for rounded crosses. Their parameter $W$ is related to $\rho$ by $W=2/\rho$. Densest packing reported in that paper reaches from about
$\approx 0.36$ for $\rho=0.1$ (our TSQ value: 0.336) over $\approx 0.55$ at $\rho= 0.2$ (our TSQ: 0.567) to about 0.92 for $\rho=0.8$ (our TSQ: 0.989).
For thin crosses, these analytical results agree very well. Naturally, the differences between the densest packing of rounded and flat crosses
increase with increasing arm width $\rho$. Therefore, our structure of flat crosses reaches $\phi_{\max}=1$ when $\rho=1$, while the rounded
crosses reach a smaller fill level at $W=2$. They leave open areas next to the rounded ends, while the crystalline packings of the flat end crosses can cover the complete area.

The result of our experiments on random packing are the following: The uniaxial compression creates an almost isotropic jammed state
of crosses, with an induced orientational order restricted essentially to the first layer at the boundaries.
The measurements of the slider positions and the related areas occupied by a fixed number of crosses provide the packing fractions at the jamming transition.
These jamming transitions occur at fill fractions $\phi(\rho)$ much smaller than the optimum crystalline packing for all cross
geometries studied.
It turns out that the geometry of the arm ends (rounded or flat) has only little influence on the random packing efficiency. Even though
one would intuitively assume that rounded ends lead to a denser packing, the effect is only in the range of a few percent.

There are a few available literature data to compare our results with. The results of numerical simulations reported by Marschall and Teitel \cite{Marschall2020} for rounded crosses with aspect ratio of $\rho = 0.5$ is in excellent agreement with our experimental findings of $0.62 < \phi< 0.66$ for the rounded crosses with $\rho = 0.445$ and $0.67 <\phi < 0.71$ for rounded crosses with $\rho=0.53$.

These pair correlation functions reported by these authors also appear very similar to our
findings in Fig. \ref{fig:order}. Marschall and Teitel presented the numerically computed orientational correlations not as a function of the cross separation $r$,
but they distinguished between neighbors in the directions along the arms and in the directions 45$^\circ$ respective to the arms.
As in our experiments, the density correlations vanish at distances larger than $\approx 3\dots 4~r_0$. The authors also computed the tetratic
correlation, also separately for neighboring crosses in the direction along the arms and diagonal to them. The features of these plots
are qualitatively very similar to our experimental data. A first positive peak $Q_4(r_0)\approx 1$ is followed by a negative peak and
thereafter alternating excursions of $Q_4$ to positive and negative peaks until the correlations vanish at around $3$ to $4~r_0$.

The stress-birefringent crosses studied by Zheng et al. \cite{Zheng2017}, practically equivalent to
rounded crosses with $\rho=0.4$, show a jamming transition after uniaxial
compression at $\phi \approx 0.475\pm 0.002$. Interestingly, this value is considerably smaller than in our experiments with similar
crosses, and also much smaller than the value reported by Marschall and Teitel \cite{Marschall2020} for comparable objects.
The reason could be a considerably higher friction coefficient (0.7) of these crosses. This observation might be a hint that a systematic study of the influence of the friction coefficient on the jamming transition of crosses is desirable.

Another interesting complexion of this problem is the introduction of restricted geometries. Our experiment already shows (Fig.~\ref{fig:images})
that orientational order is induced by the boundaries, and thus the restriction to ensembles of crosses to narrow channels may lead to
structural transitions when the gap width is gradually changed. A similar scenario has been investigated by Gurin et al. \cite{Gurin2016}
and Bautista-Carbajal et al. \cite{Bautista2018} for squares (crosses in the limit $\rho \rightarrow \infty$) in a thin slit. Depending upon
the width $H$ of the slit relative to the side length of the squares, the maximum packing fractions of the structures were determined.
They vary between 0.55 and 1, undergoing several structural transitions with increasing $H$. Coexisting structures are observed near the
transition points.

Summarizing, we have proposed regular (crystalline) structures that provide the highest packing fractions of crosses over the full parameter
range of aspect ratios from $\rho=0$ (crossed line segments) to $\infty$ (squares). Depending upon $\rho$, four qualitatively different
structures have been identified that presumably provide the best packing. The results were obtained for crosses with flat ends. For crosses with
rounded arm ends \cite{Meng2020}, the optimal structures apparently provide lower packing fractions, at least for $\rho<1$.

In experiments, we have determined the jamming transitions of crosses with aspect ratios between 0.25 and 2. The packing structures achieved by
uniaxial compression are satisfactorily isotropic in the bulk, while at the moving edge some alignment is induced. This alignment leads to a
negative tetratic order parameter at the boundary (crosses diagonal to the edge) when the aspect ratio is small, and to a slightly positive
tetratic order parameter (arms along and perpendicular to the edge) at large $\rho$. Our estimate is that the transition between the two alignment
scenarios occurs at $\rho_c=\sqrt{1/2}$.

The orientational correlation parameter $Q_4$ in the jammed configurations has been determined from the experimental images.
While crosses in direct contact ($r=r_0$) create a peak at perfect order $Q_4(r_0)=1$, this peak is followed by a dip
of negative $Q_4$ and several oscillations of $Q(r)$. These decay over a range of about $3 \dots 4 r_0$, this is about the size of 2 crosses.
Over larger distances, the correlations are lost.

Positional correlations $\phi_{\rm loc}(r)$ were also determined. We observe several local peaks which can be attributed to special arrangements
of groups of crosses. The correlations decay over distances of about $3 \dots 4 r_0$, like the orientational correlations. The average packing
fractions in the experiments have been determined as a function of $\rho$. The general trend is that we achieve roughly 70~\% of the maximum
packing fractions. Rounded crosses show a slight tendency to pack tighter.

The results of this study may serve as a guide to select parameters for further experiments with cross-shaped particles such as shear experiments
or discharge of non-convex particles from containers with narrow orifices. It may also help to judge the accuracy of simulations and provide a
basis for an estimation of the stability of packing structures.

\section*{Acknowledgments}
The authors kindly acknowledge Torsten Trittel and Christoph Klopp for 3D-printing crosses, and Eckard Specht for stimulating discussions.
The German Science Foundation is acknowledged for funding within project STA 425/46-1.


\section*{Appendix A: Other crystalline configurations}

\begin{figure*}[htbp]
\centerline{
\includegraphics[width=0.8\textwidth]{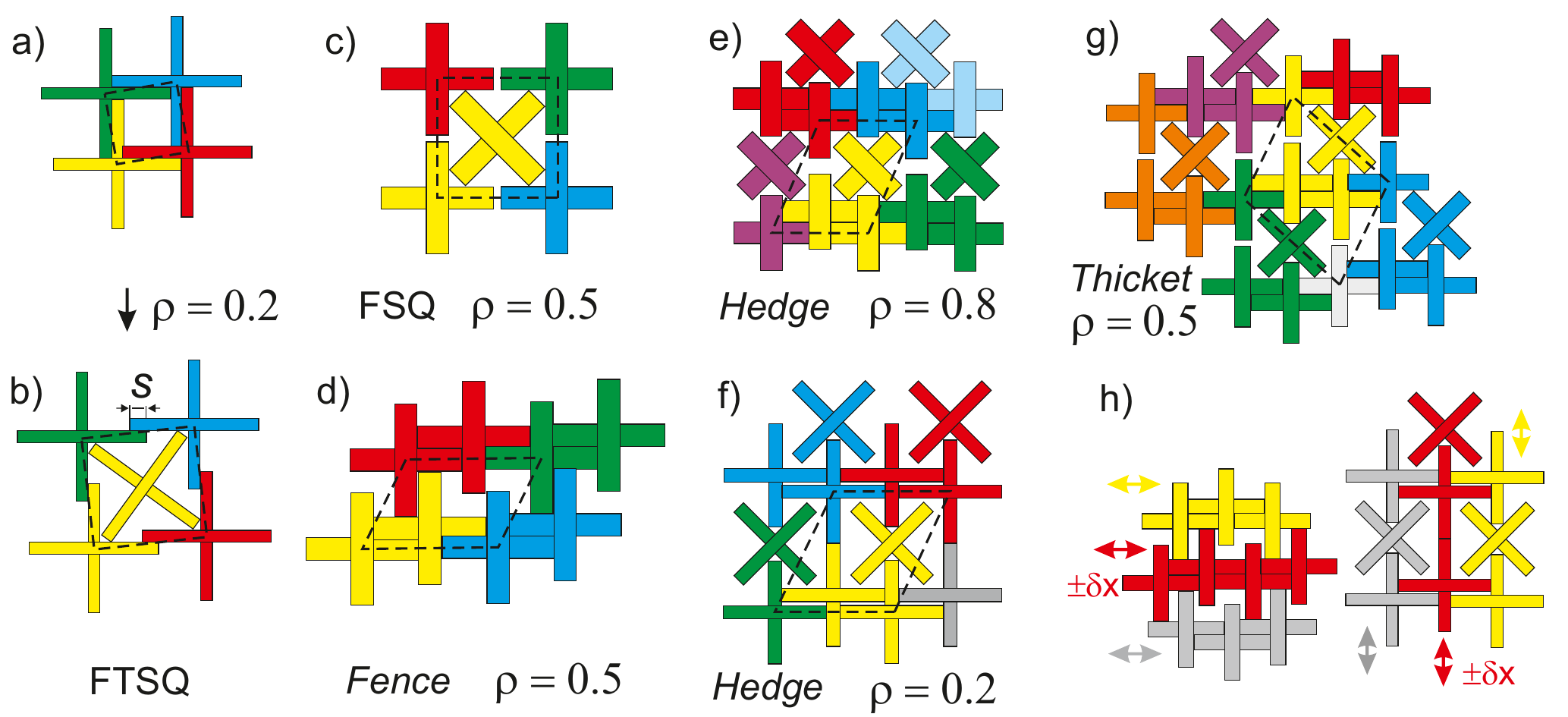}
}
\caption{\label{fig:A1}
Configurations with more than one cross per unit cell (indicated by dashed lines).
a,b) Transformation of the TSQ into a FTSQ lattice; c) filled square;
d) the {\em Fence} with two crosses per rhomboid unit cell;
e) the {\em Hedge} with three wide crosses ($\rho>2-\sqrt{2}$) per unit cell; f) same with thin crosses ($\rho<2-\sqrt{2}$;
g) more complex structure ({\em 'Thicket'}) with four crosses per unit cell.
h) demonstrates that the {\em Fence} and the thin cross {\em Hedge} in (d) and (f) are infinitely degenerate. They consist of
sub-lattices (here: equally colored crosses in subfigure h) that can be independently shifted relative to their neighbors
by small displacements $ \delta x$.
}
\end{figure*}
\begin{figure}[htbp]
\centerline{
\includegraphics[width=0.99\columnwidth]{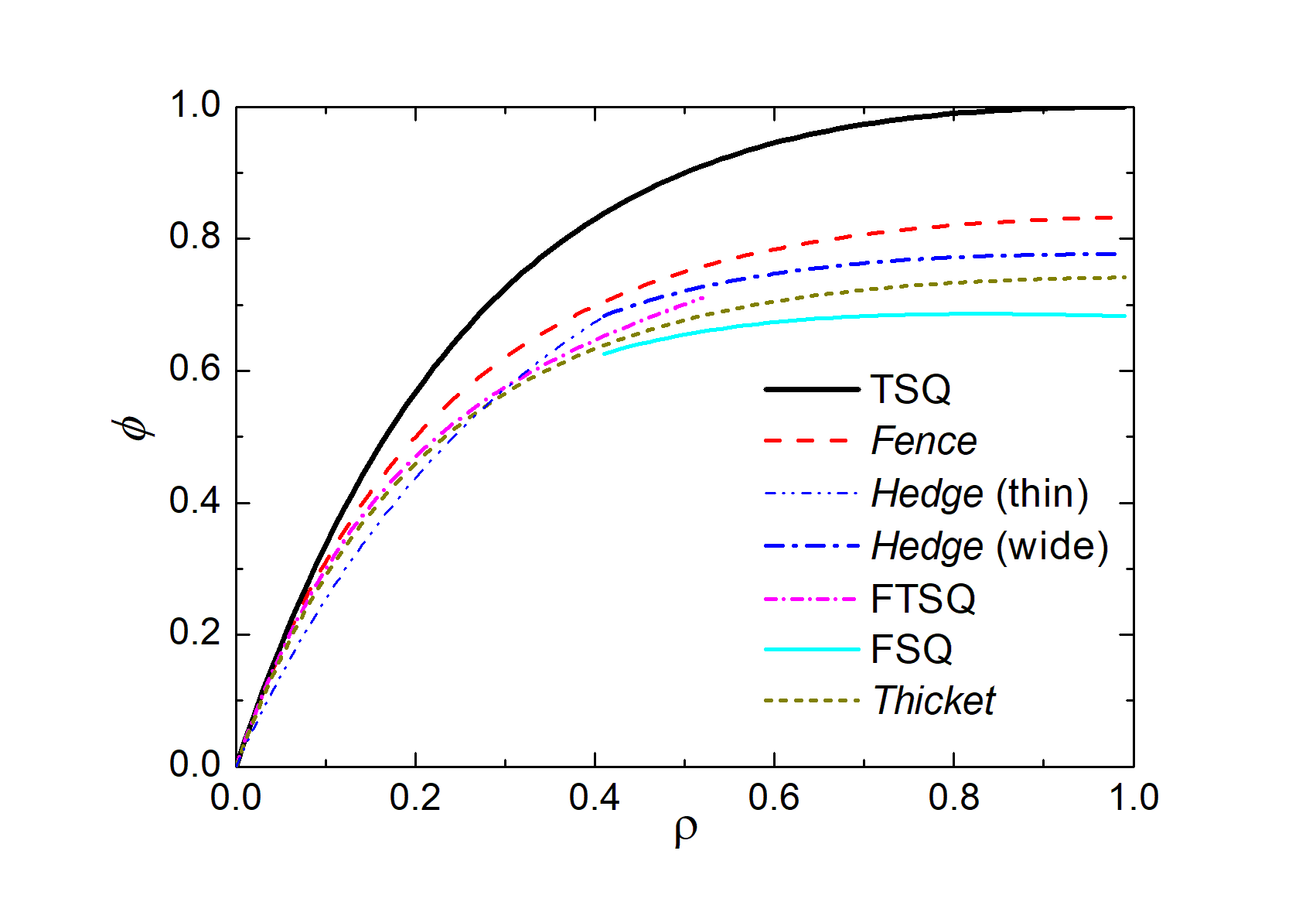}
}
\caption{
Packing fractions of selected configurations in the range $\rho<1$. The corresponding unit cells are shown in Fig.~\ref{fig:A1}.
Some of the configurations do not exist in the complete parameter range. For example, there is a crossover between the Hedge structures at
$\rho=2-\sqrt(2)$.
\label{fig:A2}
}
\end{figure}

In Sec.~\ref{sec:cryst}, our analysis was restricted to configurations with one cross per unit cell. In order to probe into other regular lattices,
For very small $\rho$, one might assume that a slight modification of the TSQ arrangement is more efficient. The unit cell is enlarged and filled, viz.
a diagonal cross is inserted in the center (Figs.~\ref{fig:A1} a,b). This creates the filled tilted square structure, FTSQ. The optimum angle $\gamma$
of the central cross respective to the four outer crosses is related to the aspect ratio by
$\sin 2\gamma = 1-\rho^2$. The outer crosses have to be shifted outward so that the arm lengths do not overlap along their full width $a$, but only
by a length $s= 2a-(b+a)\sqrt{2-b^2/a^2}$ (Fig.~\ref{fig:1}). Therefore, the unit cell has edge lengths
$r_{\rm FT}=a\sqrt{(\rho +(1+\rho)\sqrt{2-\rho^2})^2+ \rho^2}$.
Now, there are two crosses per FTSQ unit cell. It turns out that the resulting fill fraction
\begin{equation}
\phi_{\rm (FTSQ)}= \frac{2A_0}{A_{\rm FT}} = \frac{8\rho+2\rho^2}{(\rho +(1+\rho)\sqrt{2-\rho^2})^2+ \rho^2}
\end{equation}
is generally lower than that of TSQ for all aspect ratios $\rho>0$. We can discard this configuration as candidate for the optimum packing.
At $\rho \approx 0.52$, the overlap $s$ reaches zero and the configuration does not exist further.
The graph of this function is shown in Fig.~\ref{fig:A2}.

The inner cross can also be inserted symmetrically when the arms of the outer frame do not overlap, as shown in Fig.~\ref{fig:A1}c).
The area of the quadratic unit cell of this Filled Square (FSQ) is $A_{\rm FS}=[\sqrt(2)(a+b)+b]^2$, and the packing fraction,
with two crosses per unit cell, is
\begin{equation}
\phi_{\rm (FSQ)}= \frac{2A_0}{A_{\rm FS}} = \frac{8\rho+2\rho^2}{(\sqrt{2}(1+\rho)+\rho)^2}
\label{eq:FSQ}
\end{equation}
It is less efficient than FTSQ everywhere. For $\rho<\sqrt{2}-1$ in particular, the outer crosses cannot get closer since their arms forming the
sides of the unit cell touch each other. Then, its packing efficiency is even worse than predicted by an extrapolation of Eq.~(\ref{eq:FSQ}).

Another promising configuration, with all crosses in the same orientation, is the '{\em Fence}', shown in Fig.~\ref{fig:A1}c.
The rhomboid unit cell with area
$A_{\rm Fe}= [(2a+2b)(a+2b)]^2$ is sketched by dashed lines. It contains two crosses per unit cell, and
\begin{equation}
\phi_{\rm (Fence)}= \frac{2A_0}{A_{\rm Fe}} = \frac{8\rho+2\rho^2}{ (2+2\rho)^2 (1+2\rho)^2}.
\end{equation}
As is seen in Fig. \ref{fig:A2}, it indeed provides the best packing fraction of all considered structures except TSQ.

Three crosses per unit cell are contained in the structures in Fig.~\ref{fig:A1}e,f, the '{\em Hedge}'. There are two different configurations:
For thin crosses ($\rho<2-\sqrt{2}$), the diagonal cross touches the end of the vertical arms of the upper and lower neighbors, and for wide crosses,
 it touches the horizontal arms of the upper and lower crosses at their sides.  The unit cell areas are
$A_{\rm Hw} = (2a+2b)(\sqrt{2}(a+b)+2b)$ for the wide crosses ($\rho>2-\sqrt{2}$) and
$A_{\rm Ht} = (2a+2b)^2 $ for the thin crosses ($\rho<2-\sqrt{2}$). Packing fractions are shown as blue lines in Fig.~\ref{fig:A2}.
One may conceive other crystalline configurations, even such with more than three crosses per unit cell (e.~g. four in the '{\em Thicket}',
Fig. \ref{fig:A1}g).
They all fail to provide a better packing than the optimum, TSQ. For the {\em Thicket} shown in the image, for example, one finds a packing fraction
\begin{equation}
\phi_{\rm (Thicket)}= \frac{4A_0}{A_{\rm Th}} = \frac{16\rho+4\rho^2}{(1+\rho) (3\sqrt{2}(1+\rho)+5\rho)}.
\end{equation}
It is comparable to those of the FTSQ and {\em Hedge} structures.
This list is by far not complete. Generally, it seems that packing diagonal crosses into
unit cells brings no advantages over structures where all crosses are oriented uniformly.

Finally, we note a certain peculiarity of two of the structures, demonstrated in Fig.~\ref{fig:A2}h: The {\em Fence} and the thin {\em Hedge} both
possess a soft mode, they consist of one-dimensional sub-lattices. Horizontal rows of the {\em Fence} in Fig.~\ref{fig:A1}d can be shifted respective to their neighbor rows by a distance $\delta x$ of at most $a(1-\rho)$ without changing the packing fraction. The structure is therefore infinitely degenerate. The same applies to the {\em Hedge} of Fig.~\ref{fig:A1}f.
Each vertical line can be shifted with respect to its neighbors by a small displacements $|\delta x|<(2-\sqrt{2}(1+\rho))a$.
\\

\section*{Appendix B: Surface induced order}

\begin{figure}[htbp]
\centerline{
\includegraphics[width=0.66\columnwidth]{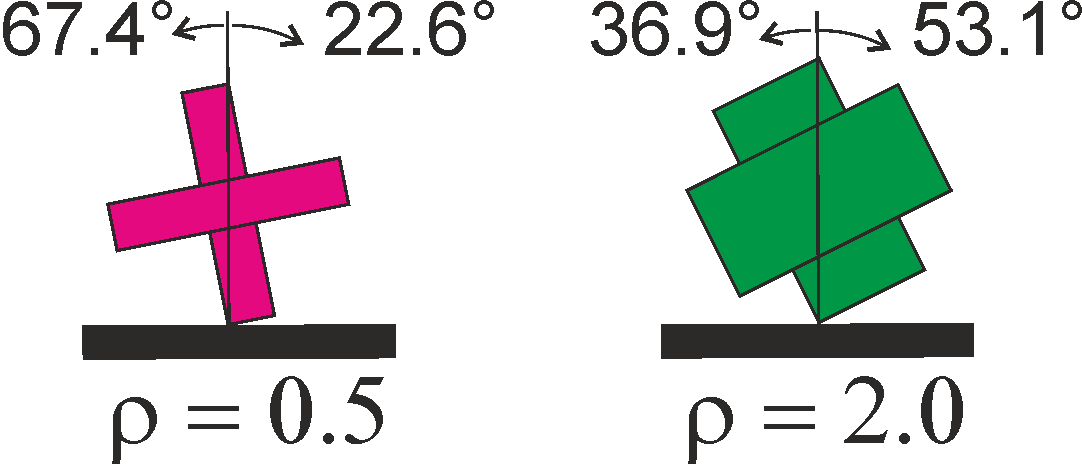}
}
\caption{\label{fig:B1}
Crosses in contact with the wall at one corner: Depending on the angle between cross and
barrier, the cross rotates either towards $\alpha_i=\pi/4$ (here counterclockwise) or $\alpha_i=0$ (here clockwise).
The probabilities relate to angular ranges $2\alpha_c$ and $\pi/2-2\alpha_c$ corresponding to the two scenarios.
Their widths are given for the two selected $\rho$.
}
\end{figure}

When a cross gets in contact with the moving barrier, a torque is generated by the force of the moving barrier acting at the contact point
and the inertia force acting on the center of mass.
Depending on the sign of this torque, the cross has two possible reorientations, either towards $\alpha_i=0$
or towards $\alpha_i=\pi/4$ (Fig.~\ref{fig:B1}). The critical angle separating these two cases is $\alpha_c= \pi/2-\arctan (2\rho+1)$.
The probabilities are equal for $\alpha_c=\pi/8$ ($\rho=\sqrt{1/2}$). Thicker crosses prefer alignment parallel to the walls, thinner ones
prefer diagonal surface alignment.

\section*{Appendix C: Experimental parameters}

\begin{table}[h!]
\centering
{
		\renewcommand{\arraystretch}{1.5}
		\begin{tabular}  {|r |c| c| c|}
			
			\hline
			 nominal &aspect ratio $\rho$      & no. of crosses   & no. of exp. \\
			\hline
			3 mm  &\,0.248  &  401  & 100  \\
			
			2 mm   &0.284  &  500  & 100  \\
			
			4 mm  &0.445 &  400   & 50     \\

			5 mm  &0.516 &  400    & 100 \\
			
			8 mm  &0.728 &  312   & 100\\
			
            10 mm &0.894 &  191  & 100\\

            black & 2.0& 166 & 50\\
        \hline
  		\end{tabular}
	}

	\caption{Minimal number of crosses used in each experiment and number of repetitions of the
experiments. In each individual experiment, three images are recorded.
}
	\label{tab:2}%
\end{table}

Table \ref{tab:2} lists the number of experiments and the number of crosses for each type of crosses.

\end{document}